\documentstyle[11pt,amsmath,amssymb,epsf,epic]{article}

\numberwithin{equation}{section}

\renewcommand{\i}{\ri}

\newcommand{\fa}{{\frak A}}
\newcommand{\fm}{{\frak M}}
\newcommand{\fh}{{\frak h}}

\newcommand{\ra}{\rightarrow}
\newcommand{\tr}{{\rm Tr}}

\newcommand{\bra}{\langle}
\newcommand{\ket}{\rangle}

\newcommand{\be}{\begin{equation}}
\newcommand{\ee}{\end{equation}}
\newcommand{\bea}{\begin{eqnarray}}
\newcommand{\eea}{\end{eqnarray}}

\newcommand{\ce}{{\cal E}}
\newcommand{\ch}{{\cal H}}
\newcommand{\cc}{{\cal C}}

\newcommand{\cs}{{\cal S}}

\newcommand{\mqn}{M_{Q_n}}
\newcommand{\e}{{\rm e}}
\newcommand{\p}{{\rm p}}
\renewcommand{\d}{{\rm d}}
\newcommand{\tbeta}{\beta'}
\newcommand{\ext}{{\rm ext}}

\newcommand{\betak}{\beta_{{\cal E}_k}}

\newcommand{\grintl}{[\kern-.18em [}
\newcommand{\grintr}{]\kern-.18em ]}
\newcommand{\ds}{\displaystyle}

\newcommand{\sinc}{{\rm sinc}}

\newcommand{\f}{{\rm f}}

\newcommand{\h}{\cH}

\newcounter{resultcounter}[section]

\newtheorem{thm}[resultcounter]{Theorem}
\newtheorem{lem}[resultcounter]{Lemma}
\newtheorem{prop}[resultcounter]{Proposition}

\newtheorem{definition}[resultcounter]{Definition}
\newtheorem{rem}[resultcounter]{Remark}

\def\bed{\begin{definition}}
\def\eed{\end{definition}}

\def\one{{\mathchoice {\rm 1\mskip-4mu l} {\rm 1\mskip-4mu l} {\rm 1\mskip-4.5mu l} {\rm 1\mskip-5mu l}}}

\def\proof{\noindent{\bf Proof.}\ \ }

 \def\cB{{\cal B}} \def\cC{{\cal C}}
 \def\cE{{\cal E}} \def\cF{{\cal F}}
 \def\cH{{\cal H}} 
  
\def\cM{{\cal M}}  
  
\def\cS{{\cal S}}

\newcommand{\R}{{\mathbb R}}
\newcommand{\N}{{\mathbb N}}

\newcommand{\C}{{\mathbb C}}

\newcommand{\E}{{\mathbb E}}
\renewcommand{\P}{{\mathbb P}}

\newcommand{\cme}{{\cal M}_{(E)}}

\def\proof{\noindent{\bf Proof.}\ \ }
\def\qed{\hfill $\Box$\medskip}

\newcommand{\ri}{{\rm i}}
\newcommand{\fer}[1]{(\ref{#1})}
\newcommand{\scalprod}[2]{\left\langle {#1}, {#2}\right\rangle}
\newcommand{\bbbone}{\mathchoice {\rm 1\mskip-4mu l} {\rm 1\mskip-4mu l}
{\rm 1\mskip-4.5mu l} {\rm 1\mskip-5mu l}}
\newcommand{\po}{\overline\omega}

\begin{document}
\title{Random Repeated Interaction Quantum Systems}
\author{ Laurent Bruneau\footnote{
CNRS-UMR 8088 and D\'epartement de Math\'ematiques,
Universit\'e de Cergy-Pontoise, Site Saint-Martin, BP 222,
 95302 Cergy-Pontoise, France. Email: laurent.bruneau@u-cergy.fr, http://www.u-cergy.fr/bruneau
}\ , Alain Joye\footnote{
Institut Fourier, UMR 5582, CNRS-Universit\'e de Grenoble I
BP 74, 38402 Saint-Martin d'H\`eres, France. Partly supported by the {\it Minist\`ere fran\c cais des affaires \'etrang\`eres} through a {\it s\'ejour scientifique haut niveau}; Email: Alain.Joye@ujf-grenoble.fr
}\ , Marco Merkli\footnote{Department of Mathematics and Statistics, Memorial University, St. John's, NL, A1C 5S7, 
Canada. Partly supported by the {\it Minist\`ere fran\c cais des affaires \'etrang\`eres} through a {\it s\'ejour scientifique haut niveau}; Email: merkli@math.mun.ca, http://www.math.mun.ca/\
$\widetilde{}$ merkli/ } }

\date{\today}

\maketitle
\vspace{-1cm}
\begin{abstract}
We consider a quantum system $\cS$ interacting sequentially with independent systems $\cE_m$, $m=1,2,\ldots$ Before interacting, each $\cE_m$ is in a possibly random state, and each interaction is characterized by an interaction time and an interaction operator, both possibly random. We prove that any initial state converges to an asymptotic state almost surely in the ergodic mean, provided the couplings satisfy a mild effectiveness condition. We analyze the macroscopic properties of the asymptotic state and show that it satisfies a second law of thermodynamics.

We solve exactly a model in which $\cS$ and all the $\cE_m$ are spins: we find the exact asymptotic state, in case the interaction time, the temperature, and the excitation energies of the $\cE_m$ vary randomly. We analyze a model in which $\cS$ is a spin and the $\cE_m$ are thermal fermion baths and obtain the asymptotic state by rigorous perturbation theory, for random interaction times varying slightly around a fixed mean, and for small values of a coupling constant.
\end{abstract}

\thispagestyle{empty}
\setcounter{page}{1}
\setcounter{section}{1}

\setcounter{section}{0}

\section{Introduction}\label{sec:intro}

This paper is a contribution to rigorous non-equilibrium quantum statistical mechanics, examining the asymptotic properties of random repeated interaction systems. The paradigm of a repeated interaction system is a cavity containing the quantized electromagnetic field, through which an atom beam is shot in such a way that only a single atom is present in the cavity at all times. Such systems are fundamental in the experimental and theoretical investigation of basic processes of interaction between matter and radiation, and they are of practical importance in quantum optics and quantum state engineering \cite{MWM,WVHW,WBKM}.

A repeated interaction system is described by a ``small'' quantum system $\cS$ (cavity) interacting 
successively with independent quantum systems
$\cE_1$, $\cE_2, \ldots$ (atoms). At each moment in time, $\cS$ interacts
precisely with one $\cE_m$ (with increasing index as time
increases), while the other elements in the chain
$\cC=\cE_1+\cE_2+\cdots$ evolve freely according to their
intrinsic (uncoupled) dynamics. The complete evolution
is described by the intrinsic dynamics of $\cS$ and of $\cE_m$, plus an interaction between $\cS$ and $\cE_m$, for each
$m$. The latter consists of an interaction time $\tau_m>0$, and an
interaction operator $V_m$ (acting on $\cS$ and $\cE_m$); during
the time interval $[\tau_1+\cdots+\tau_{m-1},
\tau_1+\cdots+\tau_{m})$, $\cS$ is coupled to $\cE_m$ via a
coupling operator $V_m$. One may view $\cC$ as a ``large system'', and hence $\cS$ as an open quantum system. From this perspective, the main interest is the effect of the coupling on the system $\cS$. Does the system approach a time-asymptotic state? If so, at what rate, and what are the macroscopic (thermodynamic) properties of the asymptotic state?  Idealized models with constant repeated interaction, where $\cE_m=\cE$, $\tau_m=\tau$, $V_m=V$, have been analyzed in \cite{bjm, WBKM}. It is shown in \cite{bjm} that the coupling drives the system to a $\tau$-periodic asymptotic state, at an exponential rate. The asymptotic state satisfies the second law of thermodynamics: energy changes are proportional to entropy changes, with ratio equal to the temperature of the chain $\cC$. In experiments, where repeated interaction systems can be realized as ``One-Atom Masers'' \cite{MWM,WVHW,WBKM}, $\cS$ represents
one or several modes of the quantized electromagnetic field in a
cavity, and the $\cE$ describe atoms injected into
the cavity, one by one, interacting with the radiation while
passing through the cavity, and then exiting. It is clear that 
neither the interaction ($\tau_m$, $V_m$), nor the state of the incoming  
elements $\cE_m$ can be considered exactly the same in each interaction step $m$. Indeed, in experiments, the atoms are ejected
from an atom oven, then cooled down before entering the cavity -- a process that cannot be controlled entirely. It is therefore natural to build a certain randomness into the description. For instance, we may consider the temperature of the incoming $\cE$ or the interaction time $\tau$ to be random. (Other parameters may vary randomly as well.) We develop in this work a theory that allows us to treat repeated interaction processes with time-dependent (piecewise constant) interactions, and in particular, with {\it random interactions}. We are
not aware of any theoretical work dealing with variable or 
random interactions, other than \cite{bjm2}. Moreover, to our knowledge, this is the only work, next to \cite{bjm2}, where {\it random} positive temperature Hamiltonians (random Liouville operators) are examined.

The purpose of the present paper is twofold:

-- {\it Firstly}, we establish a general framework for random repeated interaction systems and we prove convergence results for the dynamics. The dynamical process splits into a decaying and a flucutating part, the latter converging to an explicitly identified limit in the ergodic mean. To prove the main convergence result, Theorem \ref{thmintro1} (see also Theorems \ref{thm:bjm2} and \ref{mgrethm1}), we combine techniques of non-equilibrium
quantum statistical mechanics developed in \cite{bjm} with techniques of \cite{bjm2}, developed to analyze infinite products of random operators. We generalize results of \cite{bjm2} to time-dependent, ``instantaneous'' observables. This is necessary in order to be able to extract physically relevant information about the final state, such as energy- and entropy variations. We examine the macroscopic properties of the asymptotic state and show in Theorem \ref{thmintro4} that it satisfies a second law of thermodynamics. This law is universal in the sense that it does not depend on the particular features of the repeated interaction system, and it holds regardless of the initial state of the system.

-- {\it Secondly}, we apply the general results to concrete models where $\cS$ is a spin and the $\cE$ are either spins as well, or they are thermal fermion fields. We solve the spin-spin system exactly: Theorem \ref{thmintro5} gives the explicit form of the final state in case the interaction time, the excitation level of spins $\cE$ or the temperatures of the $\cE$ are random. The spin-fermion system is not exactly solvable. We show in Theorem \ref{mgrethm2} that, for small coupling, and for random interaction times $\tau$ and random temperatures $\beta$ of the thermal fermi fields $\cE$, the system approaches a deterministic limit state. We give in Theorem \ref{thmintro3} the explicit, rigorous expansion of the limit state for small fluctuations of $\tau$ around a given value $\tau_0$. This part of our work is based on a careful execution of rigorous perturbation theory of certain non-normal ``reduced dynamics operators'', in which random parameters as well as other, deterministic interaction parameters must be controlled simultaneously.

%%%%%%%%%%%%%%%%%%%%%%%%%%%%%%%%%%%%%%%%%%%%%%%%%%%%%%%%%%%%%%%%%%%%%%%%%%%%%%%
%%%%%%%%%%%%%%%%%%%%%%%%%%%%%%%%%%%%%%%%%%%%%%%%%%%%%%%%%%%%%%%%%%%%%%%%%%%%%%%

\subsection{Setup}
\label{sect:setup}

The purpose of this section is to explain parts of the formalism, with the aim to make our main results, presented in the next section, easily understandable.

We first present the deterministic description. 
According to the fundamental principles of quantum mechanics,
states of the systems $\cS$ and $\cE_m$ are given by normalized
vectors (or density matrices) on Hilbert spaces $\cH_\cS$ and
$\cH_{\cE_m}$, respectively. We assume that $\dim\cH_\cS<\infty$, 
while the $\cH_{\cE_m}$ may be infinite dimensional. 
Observables of $\cS$ and $\cE_m$ are bounded operators forming
{\it von Neumann algebras} $\fm_\cS\subset \cB(\cH_\cS)$ and
$\fm_{\cE_m}\subset \cB(\cH_{\cE_m})$. Observables $A_\cS\in
\fm_\cS$ and $A_{\cE_m}\in\fm_{\cE_m}$ evolve according to the
{\it Heisenberg dynamics} ${\mathbb R}\ni t\mapsto
\alpha^t_\cS(A_\cS)$ and ${\mathbb R}\ni t\mapsto
\alpha^t_{\cE_m}(A_{\cE_m})$ respectively, where $\alpha^t_\cS$
and $\alpha^t_{\cE_m}$ are $*$-automorphism groups of $\fm_\cS$ and
$\fm_{\cE_m}$, respectively, see e.g. \cite{BR}.  The Hilbert
space of the total system is the tensor product $\cH=
\cH_\cS\otimes\cH_\cC$, where $\cH_\cC=\bigotimes_{m\geq
1}\cH_{\cE_m}$ is the Hilbert space of the chain, and the
non-interacting dynamics is defined on the algebra
$\fm_{\cS}\bigotimes_{m\geq 1}\fm_{\cE_m}$ by
$\alpha_\cs^t\bigotimes_{m\geq 1}\alpha_{\cE_m}^t$. The infinite tensor product $\cH$ is taken with respect to distinguished ``reference states'' of the systems $\cS$
and $\cE_m$, represented by vectors $\psi_{\cS}\in\cH_\cS$ and
$\psi_{\cE_m}\in \cH_{\cE_m}$\footnote{Those vectors are to be taken 
cyclic and separating for the algebras $\fm_{\cS}$ and
$\fm_{\cE_m}$, respectively \cite{BR}. Their purpose is to fix macroscopic properties of the system. However, since dim$\cH_\cS<\infty$, the vector $\psi_\cS$ does not play any significant role. In practice, it is chosen so that it makes computations as simple as possible.}. Typically, one takes
the reference states to be equilibrium (KMS) states for the
dynamics $\alpha^t_\cS$, $\alpha^t_{\cE_m}$, at inverse
temperatures $\beta_\cS$, $\beta_{\cE_m}$.

 It is useful to
consider the dynamics in the {\it Schr\"odinger picture}. For
this, we implement the dynamics via unitaries, generated by
self-adjoint operators $L_\cS$ and $L_{\cE_m}$, acting on ${\cal
B}(\cH_\cS)$ and ${\cal B}(\cH_{\cE_m})$, respectively. The
generators, called Liouville operators, are uniquely determined by
\begin{equation}
\alpha_\#^t(A) = \e^{\ri t L_\#} A_\# \e^{-\ri t L_\#},\ t\in{\mathbb R},\ \ \mbox{and $L_\#\psi_\#=0$},
\label{m1}
\end{equation}
where $\#$ stands for either $\cS$ or $\cE_m$ \footnote{The existence and
uniqueness of $L_\#$ satisfying \fer{m1} is well known under
general assumptions on the reference states $\psi_\#$ \cite{BR}.}.
In particular, \fer{m1} holds if the reference states are
equilibrium states. Let $\tau_m>0$ and
$V_m\in\fm_{\cS}\otimes\fm_{\cE_m}$ be the interaction time and
interaction operator associated to $\cS$ and $\cE_m$. We define
the (discrete) repeated interaction Schr\"odinger dynamics of a
state vector $\psi\in\cH$, for $m\geq 0$, by
\begin{equation}
U(m)\psi = \e^{-\ri \tau_m \widetilde L_m}\cdots\e^{-\ri \tau_2\widetilde L_2}\e^{-\ri \tau_1\widetilde L_1}\psi,
\label{m2}
\end{equation}
where
\begin{equation}
\widetilde L_k = L_k+\sum_{n\neq k} L_{\cE_n}
\label{m3}
\end{equation}
describes the dynamics of the system during the time interval
$[\tau_1+\cdots+\tau_{k-1},\tau_1+\cdots+\tau_k)$, which
corresponds to the time step $k$ of the discrete process, with 
\begin{equation}
L_k=L_\cS +L_{\cE_k} + V_k,
\label{m4}
\end{equation}
acting on $\cH_\cS\otimes\cH_{\cE_k}$. (We understand that
the operator $L_{\cE_n}$ in \fer{m3} acts nontrivially only on the
$n$-th factor of the chain Hilbert space $\cH_{\cC}$.)

An operator $\rho$ on $\cH$ which is self-adjoint, non-negative, and has unit trace is called a density matrix. A state $\varrho(\cdot)={\rm Tr}(\rho\, \cdot\,)$, where $\rm Tr$ is the trace over
$\cH$, is called a {\it normal state}. Our goal is to understand
the large-time asymptotics ($m\rightarrow \infty$) of expectations
\begin{equation}
\varrho\left(U(m)^* O U(m)\right)\equiv\varrho(\alpha^m(O)),
\label{m5}
\end{equation}
for normal states $\varrho$ and certain observables $O$. Important physical observables are represented by operators that act either just on $\cS$ or ones that describe exchange processes between $\cS$ and the chain $\cC$. The latter are represented by time-dependent operators because they act on $\cS$ and, at step $m$, on the element $\cE_m$ which is in contact with $\cS$. We define {\it instantaneous} observables to be those of the form 
\begin{equation}
O= A_\cS\otimes_{j=-l}^rB_m^{(j)},
\label{intro0'}
\end{equation}
where $A_\cS\in\fm_\cS$ and $B_m^{(j)} \in \fm_{\cE_{m+j}}$ (we do not write identity operators in the tensor product). The class of instantaneous observables allows us to study all properties of $\cS$ alone, as well as exchange properties between $\cS$ and $\cC$.

Let us illustrate our strategy to analyze \fer{m5} for the initial state determined by the vector $\psi_0 = \psi_\cS\otimes\psi_\cC$, where $\psi_\cC=\otimes_{m\geq 1}\psi_{\cE_m}$. We use ideas stemming from the algebraic approach to quantum dynamical systems far from equilibrium to obtain the following representation for large $m$ (Proposition \ref{prop:instobsreduc})
\begin{equation}
\scalprod{\psi_0}{\alpha^m(O)\psi_0} = \scalprod{\psi_0}{PM_1 \cdots M_{m-l-1} N_m(O)P\psi_0}.
\label{intro1'}
\end{equation}
Here, $P$ is the orthogonal projection onto $\cH_\cS$, along $\psi_\cC$, projecting out the degrees of freedom of $\cC$. The $M_k$ are {\it effective operators} which act on $\cH_\cS$ only, encoding the effects of the interactions on the system $\cS$. They are called {\it reduced dynamics operators} (RDO), and have the form
$$
M_k = P \e^{\i \tau_k K_k}P,
$$
where $K_k$ is an (unbounded, non-normal) operator acting on $\cH_\cS\otimes\cH_{\cE_k}$, satisfying $\e^{\i t K_k} A\e^{-\i tK_k}=\e^{\i t L_k} A\e^{-\i t L_k}$ for all $A\in\fm_{\cS}\otimes\fm_{\cE_k}$, and $K_k\psi_\cS\otimes\psi_{\cE_k}=0$.\footnote{These are the defining properties of $K_k$; $K_k$ has an explicit form expressible in terms of the modular data of $(\fm_{\cS}\otimes\fm_{\cE_k},\psi_\cS\otimes\psi_{\cE_k})$, see Section \ref{ssec:initreduc}.}
The operator $N_m(O)$ acts on $\cH_\cS$ and has the expression (Proposition \ref{mgreprop})
\begin{equation}
N_m(O)\psi_0 = P \e^{\ri\tau_{m-l} \widetilde L_{m-l}}\cdots
\e^{\ri \tau_m \widetilde L_m} (A_\cS\otimes_{j=-l}^{r} B^{(j)}_m)
\e^{-\ri\tau_m \widetilde L_m}\cdots \e^{-\ri \tau_{m-l}
\widetilde L_{m-l}} \psi_0.
\label{nmo'}
\end{equation}
The asymptotics $m\rightarrow\infty$ of \fer{intro1'} for identical matrices $M_k\equiv M$ has been studied in \cite{bjm}. In the present work we consider the $M_k$ to be random operators. We allow for randomness through random interactions (interaction times, interaction operators) as well as random initial states of the $\cE_m$ (random temperatures, energy spectra, etc).

Let $(\Omega,\cF,\p)$ be a probability space. To describe the stochastic dynamic process at hand, we introduce the standard probability measure $\d\P$ on
$\Omega_\ext:=\Omega^{\N^*}$,
\be
\d\P =\Pi_{j\geq 1}\d \p_j, \ \ \ \mbox{where } \ \ \ \d \p_j\equiv \d
\p, \ \  \forall j\in \N^*.\label{intro10'}
\ee 
We make the following randomness assumptions: 
\begin{itemize}
\item[(R1)] The reduced dynamics operators $M_k$ are independent, identically distributed (iid) random operators. We write $M_k=M(\omega_k)$, where $M:\Omega\rightarrow {\cal B}({\mathbb C}^d)$ is an operator valued random variable.
\item[(R2)] The operator $N_m(O)$ is independent of the $M_k$ with $1\leq k\leq m-l-1$, and it has the form $N(\omega_{m-l},\ldots,\omega_{m+r})$, where $N:\Omega^{r+l+1}\rightarrow  {\cal B}({\mathbb C}^d)$ is an operator valued random variable.
\end{itemize}

Since the operator $M_k$ describes the effect of the $k$-th interaction on $\cS$, assumption (R1) means that we consider iid random repeated interactions. The random variable $N$ in (R2) does not depend on the time step $m$. This is a condition on the observables, it means that the nature of the quantities measured at time $m$ are the same. For instance, the $B^{(j)}_m$ in \fer{intro0'} can represent the energy of $\cE_{m+j}$, or the part of the interaction energy $V_{m+j}$ belonging to $\cE_{m+j}$, etc. Both assumptions are verified in a wide variety of physical systems: we may take random interaction times $\tau_k=\tau(\omega_k)$, random coupling operators $V_k=V(\omega_k)$, random energy levels of the $\cE_k$ encoded in $L_{\cE_k}=L_\cE(\omega_k)$, random temperatures $\beta_{\cE_k}=\beta_\cE(\omega_k)$ of the initial states of $\cE_k$, and so on; see Sections \ref{sec:spin} and \ref{sec:spinfermion} for concrete models.

%%%%%%%%%%%%%%%%%%%%%%%%%%%%%%%%%%%%%%%%%%%%%%%%%%%%%%%%%%%%%%%%%%%%%%%%%%%%%%%
%%%%%%%%%%%%%%%%%%%%%%%%%%%%%%%%%%%%%%%%%%%%%%%%%%%%%%%%%%%%%%%%%%%%%%%%%%%%%%%

\subsection{Main results}

Our main results are: the existence and identification of the limit of infinite products of random reduced dynamics operators; the proof of the approach of a random repeated interaction system to an asymptotic state, together with its identification; the analysis of the macroscopic properties of the asymptotic state; explicit expressions of that state for spin-spin and spin-fermion systems. We present here some main results and refer to subsequent sections for more information and for proofs.

{\it -- Ergodic limit of infinite products of random operators.\ } 
The asymptotics of the dynamics \fer{intro1'}, in the random case, is encoded in the product 
$$
M(\omega_1)\cdots M(\omega_{m-l-1})N(\omega_{m-l},\ldots,\omega_{m+r}).
$$
It is not hard to see that the spectrum of the operators $M(\omega)$ is contained inside the closed complex unit disk, and that $M(\omega)\psi_\cS=\psi_\cS$ (see Lemma \ref{contraction}). 
%%%%%%%%%%%%%%%%%%%%%% DEF: \cme %%%%%%%%%%%%%%%%%%%%%%%%%%%%%
\begin{definition}
\label{mgre10}
Let $\cme$ denote the set of reduced dynamics operators whose spectrum on the complex unit circle consists only of a simple eigenvalue $1$.
\end{definition}
%%%%%%%%%%%%%%%%%%%%%%%%%%%%%%%%%%%%%%%%%%%%%%%%%%%%%%%%%%%%%%
The following is our main result on convergence of products of random reduced dynamics operators (see also Theorem \ref{mgrethm1}). We denote by ${\mathbb E}[M]$ the expectation of $M(\omega)$.

\begin{thm}[Ergodic limit of infinite operator product] \ \\
\label{thmintro1} 
\!\!Suppose that $\p(M(\omega)\in\cme)\neq 0$. Then ${\mathbb E}[M]\in\cme$. Moreover, there exists a set $\widetilde{\Omega} \subset
 \Omega^{\N^*}$ of probability one 
%s.t. ${\mathbb P}(\widetilde{\Omega})=1$ and 
s.t. for any $\po=(\omega_n)_{n\in\mathbb N}\in\widetilde{\Omega}$,
\begin{equation*}
\lim_{\nu\rightarrow\infty}\frac 1\nu\sum_{n=1}^\nu M(\omega_1)\cdots
M(\omega_n) N(\omega_{n+1},\ldots,\omega_{n+l+r+1})= |\psi_\cS\ket\bra \theta|\ \E[N],
\end{equation*}
where $\theta = P^*_{1,{\mathbb E}[M]}\psi_\cS$, $P_{1,X}$ is the (Riesz) spectral projection of $X$ associated to the eigenvalue $1$, and $*$ denotes the adjoint.
\end{thm}

{\it -- Asymptotic state of random repeated interaction systems.\ }
We use the result of Theorem \ref{thmintro1} in \fer{intro1'}, where we replace $\alpha^m$ by the random dynamics, denoted $\alpha^m_{\po}$. It follows that the ergodic limit of \fer{intro1'} is $\varrho_+({\mathbb E}[N])$, where 
\begin{equation}
\varrho_+(A):= \scalprod{\theta}{A\psi_\cS},\ \ A\in\fm_\cS.
\label{intro3'}
\end{equation}
A density argument using the cyclicity of the reference state $\psi_0$ extends the argument leading to \fer{intro1'} to all normal initial states $\varrho$ on $\fm$.

\begin{thm}[Asymptotic State]
\label{thmintro2} 
Suppose that $\p(M(\omega)\in\cme)\neq 0$. There exists a set $\widetilde{\Omega}\subset\Omega^{\N^*}$ of probability one 
%s.t. ${\mathbb P}(\widetilde{\Omega})=1$ and 
s.t. for any $\po\in\widetilde{\Omega}$, for any instantaneous observable $O$, \fer{intro0'}, and for any normal initial state $\varrho$, we have 
\begin{equation}
\lim_{\mu\to\infty} \frac{1}{\mu} \sum_{m=1}^\mu \varrho \big( \alpha^m_{\po} (O)\big) =\varrho_+\big({\mathbb E}[N]\big).
\label{intro-1'}
\end{equation}
\end{thm}

{\it -- Macroscopic properties of the asymptotic state.\ } Since we deal with open systems, it is generally not meaningful to speak about the total energy (which is typically infinite). However, variations (fluxes) in total energy are often well defined. Using an argument of \cite{bjm} (see also \cite{B} for a heuristic argument based on the hamiltonian approach) one shows that the formal expression for the total energy is constant during all time-intervals $[\tau_{m-1},\tau_m)$, and that it undergoes a jump 
\be\label{def:energy}
j(m,\po):=\alpha^m_{\po}(V(\omega_{m+1})-V(\omega_m))
\ee
at time step $m$. 
The variation of the total energy between the instants $0$ and $m$ is then $\Delta E(m,\po)=\sum_{k=1}^m j(k,\po)$. The relative entropy of $\varrho$ with respect to $\varrho_0$, two normal states on $\fm$, is denoted by ${\rm Ent}(\varrho|\varrho_0)$. Our definition of relative entropy differs from that given in \cite{BR} by a sign, so that in our case, ${\rm Ent}(\varrho|\varrho_0)\geq 0$. For a thermodynamic interpretation of entropy and its relation
to energy, we assume for the next result that $\psi_\cS$
is a $(\beta_\cS,\alpha_\cS^t)$--KMS state on $\fm_\cS$, and that
the $\psi_{\cE_m}$ are $(\beta_{\cE_m}, \alpha_{\cE_m}^t)$--KMS
state on $\fm_{\cE_m}$, where $\beta_\cS$ is the inverse temperature of $\cS$, and $\beta_{\cE_m}$ are
random inverse temperatures of the $\cE_m$. 
Let $\varrho_0$ be the state on $\fm$
determined by the vector
$\psi_\cS\otimes\psi_\cc=\psi_\cS\bigotimes_m \psi_{\cE_m}$.
The change of relative entropy is denoted $\Delta S(m,\po) := {\rm Ent} (\varrho\circ\alpha^m| \varrho_0)-{\rm Ent} (\varrho|
\varrho_0)$.

\begin{thm}[Energy and entropy productions, $2^{\rm nd}$ law of thermodyn\-amics] 
\label{thmintro4} 
Let $\varrho$ be a normal state on
  $\fm$. Then
\begin{eqnarray*}
\lim_{m\to\infty} \varrho\left(\frac{\Delta E(m,\po)}{m}\right) &=:& \d E_+ = \varrho_+ \Big(
  \E\big[ P( L_\cS+V-\e^{\i\tau L} (L_\cS+V)\e^{-\i\tau L})P\big]
\Big)\ \ a.s.\\
\lim_{m\to\infty} \frac{\Delta S(m,\po)}{m} &=:& \d S_+ =\varrho_+ \Big(
  \E\big[\beta_\cE \, P( L_\cS+V -\e^{\i\tau L}(L_\cS+V)\e^{-\i\tau L}) P\big] \Big)\ \ a.s.
\end{eqnarray*}
We call $\d E_+$ and $\d S_+$ the asymptotic energy- and entropy productions; they are independent of the initial state $\varrho$. If $\beta_\cE$ is deterministic, i.e., $\po$-independent, then the system satisfies the second law of thermodynamics: $\d S_+ = \beta_\ce \d E_+$.
\end{thm}

{\it -- Explicit expressions for asymptotic states.\ } We apply our general results to spin-spin and spin-fermion systems, presenting here a selection of results, and referring the reader to Sections \ref{sec:spin} and \ref{sec:spinfermion} for additional results and more detail.

\medskip
\noindent
{\it Spin-spin systems.\ } Both $\cS$ and $\cE$
are two-level atoms with hamiltonians $h_\cS$, $h_\cE$ having ground state energy zero, and excited energies $E_\cS$ and $E_{\cE}$, repectively. The hamiltonian describing the interaction of $\cS$ with one $\cE$ is given by $h = h_\cS+ h_\cE+\lambda v$, where $\lambda$ is a coupling parameter, and $v$ induces energy exchange processes, 
\begin{equation}
v :=a_\cS\otimes a_\cE^*+a_\cS^*\otimes a_\cE.
\label{eqnv}
\end{equation}
Here, $a_{\#}$ denotes the annihilation operators and $a^*_{\#}$ the creation operators of $\#=\cS,\cE$. The Gibbs state at inverse temperature $\beta$ is given by
\begin{equation}
\varrho_{\beta,\#}(A)=\frac{\tr(\e^{-\beta h_\#}A)}{Z_{\beta,\#}}, \ \ {\rm where} \ \   Z_{\beta,\#}=\tr(\e^{-\beta h_\#}).
\label{mm110}
\end{equation}
We take the reference state to be $\psi_0=\psi_\cS\otimes_{m\geq 1}\psi_{\cE_m,\beta_m}$, where $\psi_\cS$ is the tracial state on $\cS$, and $\psi_{\cE_m,\beta_{\cE_m}}$ is the Gibbs state of $\cE_m$ (represented by a single vector in an appropriate ``GNS'' Hilbert space, see Section \ref{sec:spin}).

The following results deals with three situations: {\it 1.} The interaction time $\tau$ is random. It is physically reasonable to assume that $\tau(\omega)$ varies within an interval of uncertainty, since it cannot be controlled exactly in experiments. {\it 2.} The excitation energy of $\cE$ is random. This situation occurs if various kinds of atoms are injected into the cavity, or if some impurity atoms enter it. {\it 3.} The temperature of the incoming atoms is random. This is physically reasonable since the incoming atom beam's temperature cannot be controlled exactly in experiments.

\begin{thm}[Random spin-spin system] Set $T:=\frac{2\pi}{\sqrt{(E_\cS-E_\cE)^2+4\lambda^2}}$.
\label{thmintro5}
\begin{itemize}
\item[1.] {\rm Random interaction time.\ } Suppose that $\beta_{\cE_m}=\beta$ is constant, and that $\tau(\omega)>0$ is a random variable satisfying 
$\p \left(\tau\notin T\N \right)\neq 0$. 
Then there exists a set $\widetilde{\Omega}\subset\Omega^{\N^*}$ of probability one, such that for all $\po\in\widetilde{\Omega}$, for all normal states $\varrho$ on $\fm$ and for all observables $A$ of $\cS$,
\begin{equation}
\lim_{\mu\to\infty} \frac{1}{\mu} \sum_{m=1}^\mu \varrho(\alpha_{\po}^m(A))=\varrho_{\beta',\cS}(A),
\label{mmm01}
\end{equation}
with $\beta'=\beta_1:=\beta E_\cE/E_\cS $.

\item[2.] {\rm Random excitation energy of $\cE$.\ } Suppose that $\tau$ and $\beta_{\cE_m}=\beta$ are constant, and that $E_\cE(\omega) >0$ is a random variable satisfying $\p \left(\tau\notin T\N \right)\neq 0$. (Here, $T=T(\omega)$ is random via $E_\cE(\omega)$.) Then there exists a set $\widetilde{\Omega} \subset\Omega^{\N^*}$ of probability one s.t. for all $\po\in\widetilde{\Omega}$, for all normal initial states $\varrho$ on $\fm$ and for all observables $A$ of $\cS$, \fer{mmm01} holds with $\beta'= \beta_2:=-E_\cS^{-1} \log \big(2 \big\{1-(1-\E[e_0])^{-1}\E\big[(1-e_0)(1-2 Z_{\beta E_\cE/E_\cS,\cS}^{-1})\big]\big\}^{-1}-1 \big)$, and where
\begin{equation}
e_0 = \left|\frac{\left(E_\cS-E_\cE-\sqrt{(E_\cS-E_\cE)^2+4\lambda^2}\right)^2+4\lambda^2 e^{\ri\tau \sqrt{(E_\cS-E_\cE)^2+4\lambda^2}}}{\left(E_\cS-E_\cE-\sqrt{(E_\cS-E_\cE)^2+4\lambda^2}\right)^2+4\lambda^2}\right|^2.
\label{enot}
\end{equation}

\item[3.] {\rm Random temperature of $\cE$.\ } Suppose that $\beta(\omega)$ is a random variable, and that $\tau>0$ satisfies $\tau\notin T\N$. Then there exists a set $\widetilde{\Omega} \subset\Omega^{\N^*}$ of probability one s.t. for all $\po\in\widetilde{\Omega}$, for all normal initial states $\varrho$ on $\fm$ and for all observables $A$ of $\cS$, \fer{mmm01} holds with $\beta'=\beta_3:=-E_\cS^{-1}\log \big(\E[Z_{\beta(\omega)E_\cE/E_\cS,\cS}^{-1}]^{-1}-1\big)$.
\end{itemize}
\end{thm}

\noindent {\it Remarks.\ } 1. In the situation of point 1. of Theorem \ref{thmintro5}, we obtain the following sharper result than \fer{mmm01}. There is are constant $C,\alpha>0$, and there is a random variable $n_0(\po)$ satisfying ${\mathbb E}[\e^{\alpha n_0}]<\infty$ such that, for each $\po\in\widetilde\Omega$:  $\big| \varrho(\alpha_{\po}^n(A))-\varrho_{\tbeta,\cS}(A)\big|\leq C \e^{-\alpha n}$, for all $n\geq n_0(\po)$, all observables $A$ and all normal initial states $\varrho$.

2. If $E_\cS=E_\cE$ 
%(which is a natural condition to allow for exchange processes without energy changes), 
then $\beta_1=\beta$. In the case of identical interactions (no randomness), the system $\cS$ is therefore ``thermalized'' by the elements of the chain, a fact which was already noticed in \cite{AJ}. One might expect that for a randomly fluctuating temperature of the $\cE$, the system $\cS$ would be thermalized at asymptotic temperature equalling the average of the chain temperature. However, point 3. of the above theorem shows that this is not the case: the asymptotic temperature is in general not the average temperature. The random repeated interaction process induces a more complicated thermalization effect on $\cS$ than simple temperature averaging.

\medskip
\noindent
{\it Spin-fermion systems.\ } Let $\cS$ be a spin-$1/2$ system with Hilbert space of pure states ${\mathbb C}^2$, and Hamiltonian given by the Pauli matrix $\sigma_z$. We take the systems $\cE$ to be infinitely extended thermal fermi fields. They model dispersive environments. Let $a(k)$ and $a^*(k)$ denote the usual fermionic creation and annihilation operators, and let $a(f)=\int_{{\mathbb R}^3} \overline{ f}(k)a(k){\rm d}^3k$, $a^*(f)=\int_{{\mathbb R}^3} f(k)a^*(k){\rm d}^3k$, for square-integrable $f$.  We take the state $\varrho_\beta$ of $\cE$ to be the equilibrium state at inverse temperature $\beta$. It is characterized by $\varrho_\beta (a^*(f)a(f))= \scalprod{f}{(1+\e^{\beta h})^{-1}f}$, where the $h$ appearing in the scalar product is the Hamiltonian of a single fermion. We represent the one-body fermion space as $\fh=L^2(\R^+,\d\mu(r);{\frak g})$, 
where ${\frak g}$ is an auxiliary Hilbert space, and we take $h$
to be the operator of multiplication by $r\in\R^+$.\footnote{\label{fn1} For instance,
for usual non-relativistic, massive fermions, the single-particle
Hilbert space is $L^2({\mathbb R}^3,\d^3k)$ (Fourier space), and
the Hamiltonian is the multiplication by $|k|^2$. This corresponds
to ${\frak g}=L^2(S^2,\d\Sigma)$ (uniform measure on $S^2$), and
$\d\mu(r) = \frac12 r^{1/2}\d r$.}

At each interaction step, $\cS$ interacts with a fresh system $\cE$ for a duration $\tau$. The interaction induces energy exchanges between the two interacting subsystems, it is represented by the operator $\lambda V$, where $\lambda$ is a small coupling constant, and $V=\sigma_x\otimes[a^*(g)+a(g)]$. Here, $\sigma_x$ is the Pauli matrix and $g=g(k)\in L^2({\mathbb R}^3,{\rm d}^3k)$ is a {\it form factor} determining the relative strength of interaction between $\cS$ and modes of the thermal field. We consider random interaction times of the form $\tau(\omega) =\tau_0+\sigma(\omega)$, where $\tau_0$ is a fixed value, and $\sigma(\omega)\in[-\epsilon,\epsilon]$ is a random variable with small amplitude $\epsilon$.

\begin{thm}[Random spin-fermion system]
\label{thmintro3}
Assume that the form factor satisfies $\|(1+\e^{\beta h/2})g\|_{L^2({\mathbb R}^3,\,\d^3k)}<\infty$, and that $\p(\sigma(\omega)\in\frac\pi 2{\mathbb N}-\tau_0)\neq 1$. There is a constant $\lambda_0>0$ s.t. if $0<|\lambda|<\lambda_0$, then Theorem \ref{thmintro2} applies, and the asymptotic state $\varrho_+$, \fer{intro3'}, has the following expansion: for any $A\in\fm_\cS$,
\begin{equation}
\varrho_+(A) = q(\sigma) A_{00} + (1-q(\sigma))A_{11} +R_{\sigma,\lambda}(A),
\label{asstate}
\end{equation}
where $A_{ij}=\scalprod{i}{ A j}$, $i,j=0,1$ and $|0\ket$, $|1\ket$ are the eigenvectors of $\sigma_z$ with eigenvalues $\pm 1$. The remainder term satisfies $|R_{\sigma,\lambda}(A)|\leq C \|A\|(\epsilon^3+\lambda^2)$, where $C$ is independent of $\epsilon,\sigma,\lambda,A$.

The probabilities $q(\sigma)$ are given by
\begin{eqnarray*}
q(\sigma)&=& \frac{\alpha_+}{\alpha_++\alpha_-} + 2{\mathbb E}[\sigma]\frac{\alpha_-\xi_+ - \alpha_+\xi_-}{\tau_0^2(\alpha_++\alpha_-)^2} + 4({\mathbb E}[\sigma])^2 (\xi_++\xi_-)\frac{\alpha_-\xi_+-\alpha_+\xi_-}{\tau_0^4(\alpha_++\alpha_-)^3}\nonumber\\
&&\qquad +{\mathbb E}[\sigma^2] \frac{\alpha_-\eta_+ - \alpha_+\eta_-}{\tau_0^2(\alpha_++\alpha_-)^2},
\end{eqnarray*}
where, with $\sinc(x)=\sin(x)/x$, 
\begin{eqnarray}
\alpha_\pm &=& \int{\rm d}\mu(r)\frac{\|g(r)\|^2_{{\frak g}}}{1+\e^{-\beta r}} \left\{ \e^{-\beta r}{\rm sinc}^2\left[\frac{(r\mp 2)\tau_0}{2}\right] +\rm{\sinc}^2\left[\frac{(r\pm 2)\tau_0}{2}\right]\right\}\label{mmm3}\\
\xi_\pm &=& \tau_0 \int{\rm d}\mu(r)\frac{\|g(r)\|^2_{{\frak g}}}{1+\e^{-\beta r}} \left\{ \e^{-\beta r}{\rm sinc}\left[(r\mp 2)\tau_0\right] +\rm{\sinc}\left[(r\pm 2)\tau_0\right]\right\}\nonumber\\
\eta_\pm &=& \int{\rm d}\mu(r)\frac{\|g(r)\|^2_{{\frak g}}}{1+\e^{-\beta r}} \left\{ \e^{-\beta r} \cos\left[(r\mp 2)\tau_0\right] +\cos\left[(r\pm 2)\tau_0\right]\right\}.\nonumber
\end{eqnarray}
\end{thm}

Expansion \fer{asstate} shows in particular that to lowest order in $\lambda$, the final state is diagonal in the energy basis. This is a sign of decoherence of $\cS$ due to contact with the environment $\cC$.

\bigskip
\noindent
{\bf Organization of the paper.\ } In Section \ref{sec:reduction} we cast the dynamical problem into a shape suitable for further analysis. Our main result there is Proposition \ref{prop:instobsreduc}.  Section \ref{sec:randomprod} contains the proof of Theorem \ref{thmintro1}, and in Sections \ref{sec:asympstate} and \ref{sec:energy-entropy} we present the proof of Theorems \ref{thmintro2} and \ref{thmintro4}, respectively. In sections \ref{sec:spin} and \ref{sec:spinfermion} we present the setup and main results for spin-spin and spin-fermion systems. In particular, we give the proofs of Theorems \ref{thmintro5} and \ref{thmintro3}.

%%%%%%%%%%%%%%%%%%%%%%%%%%%%%%%%%%%%%%%%%%%%%%%%%%%%%%%%%%%%%%%%%%%%%%%%%%
%%%%%%%%%%%%%%%%%%%%%%%%%%%%%%%%%%%%%%%%%%%%%%%%%%%%%%%%%%%%%%%%%%%%%%%%%%
%%%%%%%%%%%%%%%%%%%%%%%%%%%%%%%%%%%%%%%%%%%%%%%%%%%%%%%%%%%%%%%%%%%%%%%%%%
%%%%%%%%%%%%%%%%%%%%%%%%%%%%%%%%%%%%%%%%%%%%%%%%%%%%%%%%%%%%%%%%%%%%%%%%%%

\section{Repeated interactions and matrix products}
\label{sec:reduction}

In this section, we link the repeated interaction
dynamics to products of matrices. This reduction is a purely ``algebraic'' procedure and randomness plays no role here. Throughout the paper, we assume without further mentioning it, that 
\begin{itemize}
\item[(A1)] $\dim \cH_\cS=d<\infty,$ and the reference vectors
$\psi_\#$ are cyclic and separating for $\fm_\#$ ($\#=\cS$ or
$\cE_m$).
\end{itemize}
Recall that cyclicity means that $\fm_{\#}\psi_{\#}$ is dense in
$\cH_\#$, and separability means that $A_{\#}\psi_{\#}=0 \Rightarrow A_\#=0$, $\forall A_\#\in\fm_\#$, and is equivalent to $\fm_{\#}'\psi_\#$ is dense
in $\cH_\#$, where $\fm_{\#}'$ is the commutant von Neumann
algebra of $\fm_{\#}$.

%%%%%%%%%%%%%%%%%%%%%%%%%%%%%%%%%%%%%%%%%%%%%%%%%%%%%%%%%%%%%%%%%%%%%%%%%%%%%%%%%%%%
%%%%%%%%%%%%%%%%%%%%%%%%%%%%%%%%%%%%%%%%%%%%%%%%%%%%%%%%%%%%%%%%%%%%%%%%%%%%%%%%%%%%

\subsection{Splitting off the trivial dynamics}\label{ssec:rimodel}

%We consider discrete times (time steps) $m\geq 1$, corresponding
%to physical times $\tau_1+\tau_2+\cdots +\tau_m$, where the index
%$m$ labels the interaction step during which the system ${\mathcal
%S}$ interacts with the element ${\mathcal E}_m$ of the chain
%${\mathcal C}={\mathcal E}_1+{\mathcal E}_2+{\mathcal
%E}_3+\cdots$. 
We isolate the ``free part'' of the dynamics given in
\fer{m2}--\fer{m4}, i.e. that of the elements $\cE_k$ which do not
interact with $\cS$ at a given time step $m$.
%%%%%%%%%%%%%%%%% PROP: DYNAMICS DECOMPOSITION %%%%%%%%%%%%%%%%%%%%%%%%%%%%%%%%%%%%%%
\begin{prop}\label{prop:dyndecomp} For any $m$, we have
\be\label{sch} U(m)=U_m^- \ \e^{-\ri \tau_m L_m}\cdots
\e^{-\ri\tau_1 L_1}\ U_m^+, \ee where \be\label{def:u+-} U_m^-=
\exp\left[ -\ri \sum_{j=1}^m
\sum_{k=1}^{j-1}\tau_jL_{\cE_k}\right] \quad {\rm and} \quad
U_m^+= \exp\left[ -\ri \sum_{j=1}^m
\sum_{k>j}\tau_jL_{\cE_k}\right] \ee are unitary operators which
act trivially on $\cH_\cS$ and satisfy $U_m^\pm\psi_\cC=\psi_\cC,$
$\forall m\in\N^*.$
\end{prop}
%%%%%%%%%%%%%%%%%%%%%%%%%%%%%%%%%%%%%%%%%%%%%%%%%%%%%%%%%%%%%%%%%%%%%%%%%%%%%%%%%%%%%%
\proof As the interaction Liouvillean at time $m$, $L_m$, and the
free Liouvillean $L_{\cE_k}$ commute provided $k\neq m$, we can
write successively 
\be
\begin{array}{rcl}
\e^{-\ri \tau_1\widetilde{L}_1} & = & \e^{-\ri \tau_1 L_1}\ \e^{-\ri \tau_1 \sum_{k>1}L_{\cE_k}}, \\
\e^{-\ri \tau_2 \widetilde{L}_2} & = & \e^{-\ri \tau_2 L_{\cE_1}}\e^{-\ri \tau_2 L_2}\
  \e^{-\ri\tau_2\sum_{k>2}L_{\cE_k}} \\
 & \vdots & \\
\e^{-\ri\tau_m\widetilde{L}_m} & = & \e^{-\ri\tau_m\sum_{k<m}L_{\cE_k}}\
  \e^{-\ri\tau_mL_m} \ \e^{-\ri\tau_m\sum_{k>m}L_{\cE_k}},
\end{array}
\ee
and then use this decomposition in \fer{m2}.
\qed

%%%%%%%%%%%%%%%%%%%%%%%%%%%%%%%%%%%%%%%%%%%%%%%%%%%%%%%%%%%%%%%%%%%%%%%%%%%%%%%%%%%%
%%%%%%%%%%%%%%%%%%%%%%%%%%%%%%%%%%%%%%%%%%%%%%%%%%%%%%%%%%%%%%%%%%%%%%%%%%%%%%%%%%%%

\subsection{Choosing a suitable generator of dynamics}\label{ssec:initreduc}

We follow an idea developed recently in the study of open quantum
systems far from equilibrium which allows to represent the
dynamics in a suitable way \cite{jp2002,bjm,bjm2,MMS,MMS2,MSB}. Let $J_m$ and $\Delta_m$ denote the
modular conjugation and the modular operator of the pair $(
\fm_\cS\otimes \fm_{\cE_m}, \psi_\cS\otimes \psi_{\cE_m})$,
respectively. For more detail see the above references as well as \cite{BR} for a textbook exposition. Throughout this paper, we assume the following condition on the
interaction, without further mentioning it:
\begin{itemize}
\item[(A2)] $ \Delta_m^{1/2}V_m\Delta_m^{-1/2}\in \fm_\cS\otimes
\fm_{\cE_m}, \ \ \forall m\geq 1$.
\end{itemize}
We present explicit formulae for the modular conjugation and the modular operator for the spin-fermion system in Section \ref{sec:spinfermion}. 
The Liouville operator $K_m$ at time $m$ associated to the
reference state $\psi_{\cS}\otimes\psi_{\cE_m}$ is defined as
$
K_m=L_\cS+L_{\cE_m}+V_m-J_m\Delta_m^{1/2}V_m\Delta_m^{-1/2}J_m
$. 
It satisfies  $\|\e^{\pm \ri K_m}\|\leq
\exp\{\|\Delta_m^{1/2}V_m\Delta_m^{-1/2}\|\}$. (In \cite{jp2002},
such operators are called C-Liouville operators.)  The main
dynamical features of $K_m$ are the relations 
\bea
\e^{\ri tL_m}\, A\, \e^{-\ri tL_m} & = & \e^{\ri tK_m}\, A\,\e^{-\ri
tK_m}, \mbox{\ $\forall A\in
 \fm_\cS\otimes \fm_\cC, m\geq 1, t\in\R$},\label{cliouv}\\
K_m \,\psi_\cS\otimes\psi_{\cE_m} & = & 0. \label{cliouv2}
\eea
Relation \fer{cliouv} means that $K_m$ implements the same
dynamics as $L_m$. This is seen to hold by noting that
the difference $K_m-L_m=J_m\Delta_m^{1/2}V_m\Delta_m^{-1/2}J_m$
commutes with all $A\in\fm_{\cS}\otimes\fm_\cC$ (since $J{\frak M}J = {\frak M}'$, as is known
from the Tomita-Takesaki theory of von Neumann algebras, see e.g.
\cite{BR}). The advantage of using $K_m$ instead of $L_m$ is that $\e^{\ri t
K_m}$ leaves $\psi_\cS\otimes\psi_{\cE_m}$ invariant. However, while $L_m$ is self-adjoint, $K_m$ is not even normal and unbounded.

We want to examine the large time behaviour of the evolution of a
normal state $\varrho$ on $\fm$, defined by $\varrho\circ \alpha^m
$ (see \fer{m5}). Since a normal state is a convex
combination of vector states, it is not hard to see that one has to examine the large time evolution of vector states only. More
precisely, by diagonalizing the density matrix, we can write $\rho
= \sum_{j\geq 1} p_j |\phi_j\rangle\langle\phi_j|$, where $p_j\geq
0$ and $\sum_{j\geq 1}p_j=1$, and where the $\phi_j$ are
normalized vectors in $\cH$. If we can show that $\lim_{m\rightarrow\infty} \scalprod{\phi}{\alpha^m(A)\phi} =
\varrho_\phi(A)$ exists for any
normalized vector $\phi\in\cH$, then any normal state satisfies 
\be\label{eq:asympstate}
\lim_{m\rightarrow\infty} \varrho(\alpha^m(A))
=\lim_{m\rightarrow\infty}\sum_{j\geq 1} p_j\scalprod{\phi_j
}{\alpha^m(A)\phi_j} = \sum_{j\geq 1} p_j\varrho_{\phi_j}(A).
\ee
In other words, we only have to analyze vector states
$\varrho(\cdot) = \bra \phi,\, \cdot \, \phi\ket$. If the asymptotic states $\varrho_\phi$ do not depend on the vector $\phi$, i.e. $\varrho_\phi\equiv \varrho_+$, then any normal initial state $\varrho$ has asymptotic state $\varrho_+$, by \fer{eq:asympstate}. The above argument works equally well if the pointwise limit $m\rightarrow\infty$ is replaced by the ergodic limit.

Next, since, by assumption (A1), $\psi_0=\psi_\cS\otimes\psi_\cC$, where $\psi_\cC=\otimes_{m\geq 1}\psi_{\cE_m}$, is cyclic for the
commutant $\fm'$ (which is equivalent to being separating for
$\fm$), we can approximate any vector in $\cH$ arbitrarily well by vectors 
\be\label{psib} 
\phi = B' \psi_0, 
\ee 
for some
\begin{equation}\label{bpn}
B'= B_\cS'\otimes_{n=1}^N B_n'\otimes_{n>N}\one_{\cE_n}\ \in\fm',
\end{equation}
with $B_\cS'\in\fm_\cS'$, $B_n'\in\fm_{\cE_n}'$ (with vanishing
error as $N\ra \infty$; see also \cite{bjm}). Hence, we may
restrict our attention to taking the limit $m\rightarrow\infty$ of expressions 
\be\label{basic} 
\scalprod{\psi_0 }{ (B')^* \alpha^m(A) B' \psi_0} = \scalprod{ \psi_0 }{ (B')^*
B'\alpha^m(A) \psi_0}.
\ee

%%%%%%%%%%%%%%%%%%%%%%%%%%%%%%%%%%%%%%%%%%%%%%%%%%%%%%%%%%%%%%%%%%%%%%%%%%%%%%%%%%
%%%%%%%%%%%%%%%%%%%%%%%%%%%%%%%%%%%%%%%%%%%%%%%%%%%%%%%%%%%%%%%%%%%%%%%%%%%%%%%%%%

\subsection{Observables of the small system}\label{ssec:systobsreduc}

To present the essence of our arguments in an unencumbered way, we
first consider the Heisenberg evolution of observables $A_\cS\in
\fm_\cS$, and we treat more general observables in the next
section. Consider expression \fer{basic}. Using Proposition \ref{prop:dyndecomp}, we obtain
 \bea
\alpha^m(A_\cS\otimes \one_\cC) & = & U(m)^*\ (A_\cS\otimes \one_\cC)\ U(m)\\
 & = & (U_m^+)^* \ \e^{\ri \tau_1{L_1}}\cdots \e^{\ri \tau_m{L_m}} \ (A_\cS\otimes \one_\cC)\
 \e^{-\ri \tau_m{L_m}}\cdots \e^{-\ri \tau_1{L_1}}\ U_m^+,\nonumber
\eea
where we made use of the fact that $U_m^-$ acts trivially on $\cH_\cS$. Due to the properties of
the unitary $U^{+}(m)$, specified in Proposition
\ref{prop:dyndecomp}, and due to \fer{cliouv}, \fer{cliouv2}, we have 
\bea 
\alpha^m(A_\cS\otimes \one_\cC)\, \psi_0 & = & (U^{+}_m)^* \ \e^{\ri \tau_1{L_1}}\cdots \e^{\ri \tau_m{L_m}} \
  (A_\cS\otimes \one_\cC )\ \e^{-\ri \tau_m{L_m}}\cdots \e^{-\ri \tau_1{L_1}} \psi_0\nonumber\\
 & = & (U^{+}_m)^* \ \e^{\ri \tau_1{K_1}}\cdots \e^{\ri \tau_m{K_m}} \ (A_\cS\otimes \one_\cC )\psi_0. 
\eea 
Let us introduce
$P_N=\one_\cS\otimes \one_{\cE_1}\otimes \cdots
\one_{\cE_N}\otimes P_{\psi_{\cE_{N+1}}}\otimes
P_{\psi_{\cE_{N+2}}} \otimes \cdots $, where
$P_{\psi_{\cE_k}}=|\psi_{\cE_k}\ket\bra \psi_{\cE_k} |$. From the
definition of $B'$,  (\ref{bpn}), we see that $\bra \psi_0 | (B')^* B'=\bra \psi_0 | (B')^*  B'P_N$. 
Moreover, introducing
the $m$-independent unitary operator 
\begin{equation*}
\tilde U^{+}_N:=\exp\left[ -\ri \sum_{j=1}^{N-1} \sum_{k=j+1}^N \tau_j
L_{\cE_k}\right]=P_N U^{+}_m, 
\end{equation*}
we can write, for $m>N$,
\bea
\lefteqn{
\scalprod{ \psi_0 }{ (B')^*  B'\alpha^m(A_\cS\otimes\one_\cC) \psi_0} = \scalprod{ \psi_0 }{ (B')^*  B' (\tilde U^{+}_N)^*
   P_N \e^{\ri \tau_1{K_1}}\cdots \e^{\ri \tau_m{K_m}}
     (A_\cS\otimes \one_\cC ) \psi_0}}\nonumber\\
 & = & \scalprod{\psi_0 }{ (B')^*  B' (\tilde U^{+}_N)^*
  \e^{\ri \tau_1{K_1}}\cdots \e^{\ri\tau_N{K_{N}}}P_N\e^{\ri\tau_{N+1}{K_{N+1}}}
  \cdots \e^{\ri \tau_m{K_m}}
 (A_\cS\otimes \one_\cC ) \psi_0}\nonumber.
\eea
We define the projection
\begin{equation}
\label{projectionP}
P=\one_\cS\otimes |\psi_\cC\ket\bra \psi_\cC|,
\end{equation}
and observe that 
\bea 
P_N\e^{\ri\tau_{N+1}{K_{N+1}}}\cdots \e^{\ri\tau_m{K_m}}(A_\cS\otimes \one_\cC )\psi_0
 & = & P_N\e^{\ri\tau_{N+1}{K_{N+1}}}\cdots \e^{\ri \tau_m{K_m}}
   P(A_\cS\otimes \one_\cC )\psi_0 \nonumber\\
 & = & P\e^{\ri\tau_{N+1}{K_{N+1}}}\cdots \e^{\ri\tau_m{K_m}}P(A_\cS\otimes \one_\cC )\psi_0.\nonumber 
\eea 
By a simple argument using the independence of the elements $\cE_k$ of $\cC$, we show exactly as in Proposition 4.1 of \cite{bjm},  
that for any $q\geq 1$ and any distinct integers $n_1, \cdots,
n_q$, 
\be 
P \e^{\ri\tau_{n_1}{K_{n_1}}}\e^{\ri\tau_{n_2}{K_{n_2}}}\cdots \e^{\ri
\tau_{n_q}{K_{n_q}}}P=P \e^{\ri
\tau_{n_1}{K_{n_1}}}P\e^{\ri\tau_{n_2}{K_{n_2}}}P\cdots P\e^{\ri
\tau_{n_q}{K_{n_q}}}P. 
\ee
Therefore, introducing operators $M_j$ acting on $\cH_\cS$ by
\be\label{defmj} 
P \e^{\ri \tau_j{K_{j}}}P=M_j\otimes |\psi_\cC\ket\bra \psi_\cC|, \ \ \ \mbox{or} \ \ \ M_j\simeq P
\e^{\ri \tau_j{K_{j}}}P,
\ee
we have proven the following result.
%%%%%%%%%%%%%%%%%% PROP: SMALL SYST OBS REDUCTION %%%%%%%%%%%%%%%%%%%%%%%%%%%%%%%%%
\begin{prop}\label{prop:systobsreduc} Let $A_\cS\in\fm_\cS$ and
$\phi=B'\psi_0$ with $B'$ as in \fer{bpn}. Then for any $m>N$ we
have 
\bea\label{almost}
 & & \scalprod{ \phi }{ \alpha^m(A_\cS\otimes\one_\cC) \phi} \\
 & & \quad\quad= \scalprod{ \psi_0 }{ (B')^*  B' (\tilde U^{+}_N)^* \e^{\ri
\tau_1{K_1}}\cdots \e^{\ri\tau_N{K_{N}}}P M_{N+1} M_{N+2} \cdots
M_m  (A_\cS\otimes \one_\cC ) \psi_0}.\nonumber 
\eea
\end{prop}
%%%%%%%%%%%%%%%%%%%%%%%%%%%%%%%%%%%%%%%%%%%%%%%%%%%%%%%%%%%%%%%%%%%%%%%%%%%%%%%%%%%
Proposition \ref{prop:systobsreduc} shows how the large time
dynamics of a repeated interaction system  is described by products 
\be\label{prodm}
\Psi_m=M_1M_2\cdots M_m \ \ \ \mbox{on}\ \ \cH_\cS. 
\ee 
The main features of the matrices $M_j$, inherited from those of $\e^{\ri
\tau_j{K_{j}}}$, are given in the following lemma.
%%%%%%%%%%%%%%%%% LEMMA: PROPERTIES OF M_j %%%%%%%%%%%%%%%%%%%%%%%%%%%%%%%%%%%%%%%%%%%%%
\begin{lem}[\cite{bjm}, Proposition 2.1]
\label{contraction} Assuming (A1), we have $M_j\psi_\cS=\psi_\cS$, for all $j\in \N^*$. Moreover, to any $\phi\in \cH_\cS$ there corresponds a unique $A\in
\fm_\cS$ such that $\phi=A\psi_\cS$. $|||\phi|||:=\|A\|_{\cB(\cH_\cS)}$
defines a norm on $\cH_\cS$, and as operators on $\cH_\cS$ endowed with this norm, the $M_j$ are contractions for any $j\in\N^*$.
\end{lem}
%%%%%%%%%%%%%%%%%%%%%%%%%%%%%%%%%%%%%%%%%%%%%%%%%%%%%%%%%%%%%%%%%%%%%%%%%%%%%%%%%

\noindent {\it Remark:} It follows from Lemma
\ref{contraction} that the spectrum of $M_j$ lies in
the closed complex unit disk, and that $1$ is an eigenvalue of
each $M_j$ (with \emph{common} eigenvector $\psi_\cS$).

%%%%%%%%%%%%%%%%%%%%%%%%%%%%%%%%%%%%%%%%%%%%%%%%%%%%%%%%%%%%%%%%%%%%%%%%%%%%%%%%%
%%%%%%%%%%%%%%%%%%%%%%%%%%%%%%%%%%%%%%%%%%%%%%%%%%%%%%%%%%%%%%%%%%%%%%%%%%%%%%%%%

\subsection{Instantaneous observables}
\label{ssec:instobsreduc}

So far, we have only considered observables of the system $\cS$. In
this section, we extend the analysis to the more general class of
instantaneous observables, defined in \fer{intro0'}.
Those are time-dependent observables, which,
at time $m$, measure quantities of the system $\cS$ and of a finite number of
elements $\cE_k$ of the chain, namely the element interacting at
the given time-step, plus the $l$ preceding elements and the $r$ following
elements in the chain. Physically important instantaneous observables are those with indices $j=-1,0$: they appear naturally in the study of the energy exchange process between the system $\cS$ and the chain (see Section \ref{sec:energy-entropy}); they also appear in experiments where one makes a measurement on the element right after it has interacted with $\cS$ (the atom which exits the cavity) in order to get \emph{indirect} information on the state of the latter.

%{\bf Remarks.\ } 1. Among this class of instantaneous observables, physically important ones are those with indices $j=-1, 0$: they appear naturally in the study of the energy exchange process between the system $\cS$ and the chain (see Section \ref{sec:energy-entropy}); they also appear in experiments where one makes a measurement on the element right after it has interacted with $\cS$ (the atom which exits the cavity) in order to get \emph{indirect} information on the state of the latter.
%2. For a given superscript $j$, one should imagine that the observables $B^{(j)}_m$ represent the same physical quantity which is measured on the various elements, one after the other. If the elements of the chain were identical then all the $B^{(j)}_m$ would be ``identical'' as well, only differing by a shift on the chain (see \cite{bjm}). We will come back to this when dealing with the random situation (see Section \ref{ssec:probadesc}).

The Heisenberg evolution of instantaneous observables is
computed in a straightforward  way, as for observables of the form
$A_\cS\otimes \one_\cC$. We refrain from presenting all details
of the derivation and present the main steps only.
Let 
\be\label{mgre16}
\alpha_k^{m,n}(B):=\e^{\ri(\sum_{j=n}^{m}\tau_j)
L_{\cE_k}}B\e^{-\ri(\sum_{j=n}^{m}\tau_j) L_{\cE_k}}, \ \ \ n\leq m,
\ee
denote the free evolution from time $n-1$ to $m$ of an
observable $B$ acting non trivially on $\cH_{\cE_k}$ only, with
the understanding that $\alpha_{k}^{m,n}$ equals the identity for
$n>m$. With this definition and (\ref{def:u+-}), we get 
\bea
 & & (U_m^-)^*(A_\cS\otimes_{j=-l}^rB^{(j)}_m) U_m^-\\
 & & \quad\quad\quad =A_\cS\otimes\alpha_{m-l}^{m,m-l+1}(B^{(-l)}_m)\otimes \cdots
   \alpha_{m-l}^{m,m}(B^{(-1)}_m)\otimes B^{(0)}_m\otimes \cdots \otimes B^{(r)}_m\nonumber\\
 & & \quad\quad\quad =A_\cS\otimes_{j=-l}^r \alpha_{m+j}^{m,m+j+1}(B^{(j)}_m). \nonumber
\eea
Hence,
\bea
 & & \alpha^m(A_\cS\otimes_{j=-l}^rB^{(j)}_m)\\
 & & \quad\quad\quad =(U_m^+)^*\e^{\ri \tau_1 L_1}\cdots \e^{\ri
  \tau_m L_m} (A_\cS\otimes_{j=-l}^r \alpha_{m+j}^{m,m+j+1}(B^{(j)}_m)) \e^{-\ri \tau_m L_m}\cdots
  \e^{-\ri \tau_1 L_1}  U_m^+.\nonumber 
\eea 
Consider a vector state
$\scalprod{\phi}{\cdot \phi}$, where $\phi$ is given by \fer{psib}. We proceed as in the previous section to obtain 
\bea
\lefteqn{ \scalprod{\phi}{\alpha^m(A_\cS\otimes_{j=-l}^rB^{(j)}_m)\, \phi}\nonumber}\\
& = & \scalprod{B'\psi_0}{B' (\tilde U_N^+)^*\e^{\ri \tau_1 L_1}\cdots
 \e^{\ri \tau_m L_m} \left(A_\cS\otimes_{j=-l}^{r} \alpha_{m+j}^{m,m+j+1}(B^{(j)}_m)\right)
 \e^{-\ri \tau_m L_m}\cdots \e^{-\ri \tau_1 L_1} \psi_0} \nonumber\\
& = & \scalprod{B'\psi_0}{B' (\tilde U_N^+)^*P_N\e^{\ri \tau_1
 K_1}\cdots \e^{\ri \tau_m K_m} \big( A_\cS\otimes_{j=-l}^{r}
 \alpha_{m+j}^{m,m+j+1}(B^{(j)}_m)\big)\psi_0}
\label{mgre0}
\eea
The vector to the right of $(\tilde U_N^+)^*$ can be further
expanded as
\bea
\lefteqn{ \e^{\ri \tau_1 K_1}\cdots \e^{\ri \tau_N K_N}P_N\e^{\ri
  \tau_{N+1}K_{N+1}}\cdots \e^{\ri \tau_m K_m} (A_\cS\otimes_{j=-l}^{r}
  \alpha_{m+j}^{m,m+j+1}(B^{(j)}_m))\psi_0}\nonumber \\
 & = & \e^{\ri \tau_1 K_1}\cdots \e^{\ri \tau_N K_N}P \e^{\ri
  \tau_{N+1}K_{N+1}}\cdots \e^{\ri \tau_m K_m} (A_\cS\otimes_{j=-l}^{r}
  \alpha_{m+j}^{m,m+j+1}(B^{(j)}_m))\psi_0\nonumber\\
 & = & \e^{\ri \tau_1 K_1}\cdots \e^{\ri \tau_N K_N}P M_{N+1}\cdots M_{m-l-1} \times\nonumber\\
 & & \times P \e^{\ri \tau_{m-l}K_{m-l}}\cdots \e^{\ri \tau_m K_m}
  (A_\cS\otimes_{j=-l}^{r} \alpha_{m+j}^{m,m+j+1}(B^{(j)}_m))\psi_0,
\label{mgre1}
\eea
where $P$ has been defined in \fer{projectionP}, and where we have
proceeded as in the derivation of (\ref{almost}) to arrive at the
product of the matrices $M_{N+1}\cdots M_{m-l-1}$.
We now define the operator $N_m=N_m(O)$, see \fer{intro0'}, acting on $\cH_{\cS}$ by
\be
(N_m\psi_\cS)\otimes\psi_\cC := P \e^{\ri
\tau_{m-l}K_{m-l}}\cdots \e^{\ri \tau_m K_m}
(A_\cS\otimes_{j=-l}^{r} \alpha_{m+j}^{m,m+j+1}(B^{(j)}_m))\psi_0.
 \label{mgre14}
\ee
We will also denote the l.h.s. simply by $N_m\psi_0$. The
operator $N_m$ depends on the instantaneous observable,
$N_m= N_m(A_\cS,B^{(-l)}_m,\ldots,B^{(r)}_m)$. 
It can be expressed as follows.
%%%%%%%%%%%%%%%%%%% PROP: ELEGANT EXPRESSION OF N %%%%%%%%%%%%%%%%%%%%%%%%%%%%%%%%%
\begin{prop}
\label{mgreprop} Let $\alpha^{m,n}$ denote the dynamics from time
$n$ to time $m$, i.e.,
$$
\alpha^{m,n}(\cdot) = U(m,n)^*\, \cdot \, U(m,n),
$$
where $U(m,n)=U(m)U(n)^*$, and $U(m)$ is given in \fer{sch}. Then
we have
\begin{eqnarray}
\label{mgre13} N_m\psi_0 &=& P\alpha^{m,m-l-1}\big(
A_\cs\otimes_{j=-l}^r B_m^{(j)}\big)\psi_0\\
&=&P\alpha^{m,m-l-1}(A_\cS\otimes_{j=-l}^0B_m^{(j)})\psi_0\
\prod_{k=1}^r \bra \psi_{\cE_{m+k}},
B_m^{(k)}\psi_{\cE_{m+k}}\ket.\nonumber
\end{eqnarray}
\end{prop}
%%%%%%%%%%%%%%%%%%%%%%%%%%%%%%%%%%%%%%%%%%%%%%%%%%%%%%%%%%%%%%%%%%%%%%%%%%%%%%%%%%%%

\noindent {\bf Proof.\ } The second equality is clear, since the
dynamics involves only the $\cE_k$ with indices $k\leq m$. To
prove the first equality, we use the properties of the operators
$K_j$ and the definition \fer{mgre14} to see that
\begin{eqnarray}
N_m\psi_0 &=& P \e^{\ri\tau_{m-l} L_{m-l}}\cdots \e^{\ri
\tau_m L_m} (A_\cS\otimes_{j=-l}^{r}
\alpha_{m+j}^{m,m+j+1}(B^{(j)}_m)) \times \label{mgre15}\\
&&\times \e^{-\ri\tau_m L_m}\cdots \e^{-\ri \tau_{m-l} L_{m-l}}
\psi_0. \nonumber
\end{eqnarray}
Next, we write the $\alpha_{m+j}^{m,m+j+1}$ in terms of the
generators $L_{\cE_{m+j}}$, see \fer{mgre16},
$$
\alpha_{m+j}^{m,m+j+1}(\cdot) = \e^{\ri (\tau_{m+j+1}+\cdots
+\tau_m) L_{\cE_{m+j}}}\ \cdot\ \e^{-\ri (\tau_{m+j+1}+\cdots
+\tau_m) L_{\cE_{m+j}}}.
$$
Inserting this expression into \fer{mgre15} we can distribute the
generators $L_{\cE_{m+j}}$ among the propagators in \fer{mgre15},
 and we see that
 \begin{equation*}
N_m\psi_0 = P \e^{\ri\tau_{m-l} \widetilde L_{m-l}}\cdots
\e^{\ri \tau_m \widetilde L_m} (A_\cS\otimes_{j=-l}^{r} B^{(j)}_m)
\e^{-\ri\tau_m \widetilde L_m}\cdots \e^{-\ri \tau_{m-l}
\widetilde L_{m-l}} \psi_0,
\end{equation*}
where the $\widetilde L_k$, \fer{m2}, give the full dynamics. 
\qed

Finally, $N_m$ can be defined on all of $\cH_\cS$ in the following way. From Proposition
\ref{mgreprop}, it is immediate that for all observables $A'_\cS$
in the commutant $\fm'_\cS$, we can set
$
N_mA'_\cS\psi_0 :=A'_\cS N_m\psi_0.
$
Since $\fm'_{\cS}\psi_{\cS}=\cH_{\cS}$ (separability of
$\psi_{\cS}$), $N_m$ is defined on all of $\cH_{\cS}$.
We have proven the following result.
%%%%%%%%%%%%%%%%%% PROP: INSTANTANEOUS OBS REDUCTION %%%%%%%%%%%%%%%%%%%%%%%%%%%%
\begin{prop}\label{prop:instobsreduc} Let $O$ be an instantaneous observable, \fer{intro0'}, and let $\phi=B'\psi_0$ with $B'$ as in \fer{bpn}. Then we have for any $m>N+l+1$ 
\begin{equation*}
  \scalprod{\phi}{\alpha^m(O) \phi}=\scalprod{\psi_0}{(B')^*  B' (\tilde U^{+}_N)^* \e^{\ri \tau_1{K_1}}
   \cdots \e^{\ri\tau_N{K_{N}}}P M_{N+1} \cdots M_{m-l-1} N_m(O)\psi_0},\nonumber 
\end{equation*}
where the $M_j$ are defined in \fer{defmj}, and $N_m(O)$ is given in \fer{mgre14}.
\end{prop}
%%%%%%%%%%%%%%%%%%%%%%%%%%%%%%%%%%%%%%%%%%%%%%%%%%%%%%%%%%%%%%%%%%%%%%%%%%%%%%%%%%%

To understand the large time behaviour of instantaneous
observables, we study the $n\ra\infty$ asymptotics of  products 
\be \Psi_n N_{n+l+1}=
M_1M_2\cdots M_n N_{n+l+1} \ \ \mbox{on}\ \ \cH_\cS, \label{mgre6}
\ee
where $N_{n+l+1}$ involves only quantities of the systems $\cS$ and $\cE_k$, with $k=n+1,\ldots,n+l+r+1$. The numbers $l,r$ are determined by the instantaneous observable $O$ \fer{intro0'}.

%%%%%%%%%%%%%%%%%%%%%%%%%%%%%%%%%%%%%%%%%%%%%%%%%%%%%%%%%%%%%%%%%%%%%%%%%%%%%%%%%
%%%%%%%%%%%%%%%%%%%%%%%%%%%%%%%%%%%%%%%%%%%%%%%%%%%%%%%%%%%%%%%%%%%%%%%%%%%%%%%%%
%%%%%%%%%%%%%%%%%%%%%%%%%%%%%%%%%%%%%%%%%%%%%%%%%%%%%%%%%%%%%%%%%%%%%%%%%%%%%%%%%
%%%%%%%%%%%%%%%%%%%%%%%%%%%%%%%%%%%%%%%%%%%%%%%%%%%%%%%%%%%%%%%%%%%%%%%%%%%%%%%%%

\section{Proof of Theorem \ref{thmintro1}}
\label{sec:randomprod}

According to Proposition 
\ref{prop:instobsreduc}, the large time dynamics is described by products of operators of the form \fer{mgre6}, in the limit $n\rightarrow\infty$. We will use in this section our basic assumptions (R1) and (R2), saying that the $M_j$ form a set of iid random matrices, and that $N_{n+l+1}$ is a random matrix independent of the $M_j$, $j=1,\ldots, n$. In this section,
we review results of \cite{bjm2} on products of the form
$M_1\cdots M_n$, and we extend them to products of random
matrices of the form \fer{mgre6}. Our main result here is Theorem \ref{mgrethm1}.

%%%%%%%%%%%%%%%%%%%%%%%%%%%%%%%%%%%%%%%%%%%%%%%%%%%%%%%%%%%%%%%%%%%%%%%%%%%%%%%%%
%%%%%%%%%%%%%%%%%%%%%%%%%%%%%%%%%%%%%%%%%%%%%%%%%%%%%%%%%%%%%%%%%%%%%%%%%%%%%%%%%

\subsection{Decomposition of Random Reduced Dynamics Operators}\label{ssec:decomp}

Let $P_{1,j}$ denote the spectral projection of $M_j$ for the eigenvalue $1$ (c.f. Lemma \ref{contraction}) and define
\be\label{defpsi} 
\psi_j:= P^*_{1,j}\psi_\cS, \ \ \ \ P_j:=|\psi_\cS\ket\bra \psi_j|, 
\ee 
where $P^*_{1,j}$ is the adjoint operator of $P_{1,j}$. Note that $\bra
\psi_j|\psi_\cS\ket=1$ so that $P_j$ is a projection and,
moreover, $M_j^*\psi_j=\psi_j$. We introduce the following
decomposition: 
\be\label{struct} 
M_j:=P_j+Q_jM_jQ_j,  \ \ \ \mbox{with} \ \ \ Q_j=\one-P_j. 
\ee 
The following are basic properties of products of operators $M_k$.
%%%%%%%%%%%%%%%%% PROP: PSI DECOMPOSITION %%%%%%%%%%%%%%%%%%%%%%%%%%%%%%%%%%%%%%%%
\begin{prop}[\cite{bjm2}]
\label{prop:psiform} 
We define $M_{Q_j}:=Q_jM_jQ_j$. For any $n$, we have
\be\label{eq:psi}
M_1\cdots M_n=|\psi_\cS\ket\bra\theta_n|+M_{Q_1}\cdots \mqn, 
\ee
where 
\bea \theta_n & = & \psi_n+M_{Q_n}^*\psi_{n-1}+\cdots
+M_{Q_n}^*\cdots
 M_{Q_2}^* \psi_1 \label{eq:theta1}\\
 & = & M_n^*\cdots M_2^* \psi_1
\label{eq:theta2}
\eea
and where $\bra\psi_\cS,\theta_n\ket=1$. Moreover, there exists $C_0$ such that
\begin{enumerate}
\item For any $j\in\N^*$, $\|P_j\|=\|\psi_j\| \leq C_0$ and $\|Q_j\|\leq 1+C_0$. 
\item $\ds \sup\, \{\|M_{Q_{j_n}}M_{Q_{j_{n-1}}}\cdots M_{Q_{j_1}}\|, \ n\in\N^*,\,
j_k\in{\mathbb N}^*\} \leq C_0(1+C_0)$.
\item For any $n\in\N^*$, $\|\theta_n\|\leq C_0^2$.
\end{enumerate}
\end{prop}
%%%%%%%%%%%%%%%%%%%%%%%%%%%%%%%%%%%%%%%%%%%%%%%%%%%%%%%%%%%%%%%%%%%%%%%%%%%%%%%%%%%%%%

Typically, for matrices $M_k\in\cme$ (recall Definition \ref{mgre10}), we expect the first part in the decomposition
\fer{eq:psi} to be oscillatory and the
second one to be decaying.

%%%%%%%%%%%%%%%%%%%%%%%%%%%%%%%%%%%%%%%%%%%%%%%%%%%%%%%%%%%%%%%%%%%%%%%%%%%%%%%%%
%%%%%%%%%%%%%%%%%%%%%%%%%%%%%%%%%%%%%%%%%%%%%%%%%%%%%%%%%%%%%%%%%%%%%%%%%%%%%%%%%

\subsection{The probabilistic setting}\label{ssec:probadesc}

We use the notation introduced at the end of Section \ref{sect:setup}. Let us define the shift $T:\Omega_\ext\rightarrow\Omega_\ext$ by 
\be\label{def:shift} 
(T \po)_j= \omega_{j+1}, \ \ \forall \ \po =(\omega_j)_{j\in \N}\in
\Omega_\ext. 
\ee 
$T$ is an ergodic transformation of $\Omega_\ext$. The random  reduced dynamics operators are characterized by a measurable map
\be\label{def:mprocess} 
\Omega\ni \omega_1 \mapsto M(\omega_1)\in M_d({\mathbb C}), 
\ee 
where the target space is that of all
$d\times d$ matrices with complex entries, $d$ being the dimension
of $\cH_\cS$. With a slight abuse of notation, we write sometimes $M(\po)$ instead of
$M(\omega_1)$. Hence, for any subset $B\subset M_d({\mathbb C})$, $\p (M(\po)\in B)=\p(M^{-1}(B))=\int_{M^{-1}(B)}\d\p(\omega)$, 
and similarly for other random variables. According to (R1) the product \fer{prodm} is $\Psi_n(\po) :=  M(\omega_1)M(\omega_2)\cdots M(\omega_n) =  M(T^0\po)M(T^1\po)\cdots M(T^{n-1}\po)$.

In the same way as in (\ref{defpsi}), we introduce the random
variable $\psi(\omega_1)\in {\mathbb C}^d$ defined as
\be\label{defpsi2} 
\psi(\po):= P_1(\po)^*\psi_\cS, 
\ee
where $P_1(\po)$ denotes the spectral projection of
$M(\po)$ for the eigenvalue $1$, and where ${}^*$ stands for
the adjoint. We decompose 
\be\label{structrand}
M(\po):=|\psi_\cS \ket\bra \psi(\po)|+M_Q(\po)=P(\po)+M_Q(\po) 
\ee 
as in (\ref{struct}). Note that $\psi(\po)$ and $M_Q(\po)$
define {\it bona fide} random variables: $\po\mapsto
P_1(\po)$ is measurable since $\po\mapsto M(\po)$
is \cite{az}. In the next section, we will consider the process (see \fer{eq:theta1}, \fer{eq:theta2}) 
\bea 
\theta_n(\po) & = & M^*(T^{n-1}\po)M^*(T^{n-2}\po)\cdots M^*(T\po)\psi(\po)\\ 
 & = & \sum_{j=1}^n M_Q^*(\omega_n)M_Q^*(\omega_{n-1})\cdots M_Q^*(\omega_{j+1})
  \psi(\omega_j). \nonumber 
\eea 
Note that $\theta_n$ is a Markov process, since $\theta_{n+1}(\po)=M^*(\omega_{n+1})\theta_{n}(\po)$.

Finally, the operators $N_{m}=N_m(O)$, given in \fer{nmo'} and Proposition \ref{mgreprop}, have the form
\begin{equation}
N_m(O)=N(\omega_{m-l},\ldots,\omega_{m+r}) = N(T^{m-l-1}\po),
\label{ennemm}
\end{equation}
see also condition (R2).

%%%%%%%%%%%%%%%%%%%%%%%%%%%%%%%%%%%%%%%%%%%%%%%%%%%%%%%%%%%%%%%%%%%%%%%%%%%%%%%%%
%%%%%%%%%%%%%%%%%%%%%%%%%%%%%%%%%%%%%%%%%%%%%%%%%%%%%%%%%%%%%%%%%%%%%%%%%%%%%%%%%

\subsection{Convergence results for random matrix products}\label{ssec:cvg}

We have pointed out after Lemma \ref{contraction} that the spectrum
of any RDO lies inside the complex unit disk, and $1$ is an
eigenvalue (with the \emph{deterministic}, i.e., $\po$-independent, eigenvector $\psi_\cS$). The following result on the product of an iid sequence of RDO's is
the main result of \cite{bjm2}.

%%%%%%%%%%%%%%%%% THM: CONVERGENCES FOR RRDP %%%%%%%%%%%%%%%%%%%%%%%%%%%%%%%%%%%%%%%%%%%
\begin{thm}[\cite{bjm2}]
\label{thm:bjm2}
Let $M(\omega)$ be a random reduced dynamics operator. Suppose that $\p(M(\omega)\in\cme)\neq 0$. Then we have ${\mathbb E}[M]\in\cme$. Moreover, there exist a set $\Omega_1\subset\Omega^{\N^*}$ with ${\mathbb P}(\Omega_1)=1$, and constants $C,\alpha>0$, s.t. for any $\po\in\Omega_1$ there is an $n_0(\po)$ so that 
\begin{eqnarray}
\| M_{Q}(\omega_1)\cdots M_{Q}(\omega_n)\|&\leq& C\e^{-\alpha n},\mbox{\ for all $n\geq n_0(\po)$, and}\label{eq:mqdecay} \\
 \lim_{\nu\rightarrow\infty}\frac 1\nu\sum_{n=1}^\nu\theta_n(\po) &=& \theta.
\label{eq:mqdecay1}
\end{eqnarray}
Also, $n_0(\po)$ is a random variable satisfying ${\mathbb E}[\e^{\alpha n_0}]<\infty$, and 
\begin{equation}
\theta = \left( \bbbone - {\mathbb E}[M_Q]^*\right)^{-1}{\mathbb E}[\psi]=P^*_{1,{\mathbb E}[M]}{\mathbb E}[\psi]=P^*_{1,{\mathbb E}[M]}\psi_\cS.
\label{eq:thetalim}
\end{equation}
As a consequence,
\begin{equation}
\lim_{\nu\rightarrow\infty}\frac 1\nu\sum_{n=1}^\nu M(\omega_1)\cdots
M(\omega_n) =|\psi_\cS\ket\bra \theta|=P_{1,{\mathbb E}[M]}.
\label{eq:rrdpcvg}
\end{equation}
\end{thm}
%%%%%%%%%%%%%%%%%%%%%%%%%%%%%%%%%%%%%%%%%%%%%%%%%%%%%%%%%%%%%%%%%%%%%%%%%%%%%%%%%%%%%%%%%

\noindent {\bf Remark.} In the setting of Theorem \ref{thm:bjm2}, if not only $M(\omega)$, but also
$M^*(\omega)$ has a deterministic eigenvector with eigenvalue
$1$ (denoted $\psi_\cS^*$ and normalized as $\scalprod{\psi_{\cS}^*}{\psi_\cs}=1$), then $\theta=\psi_\cS^*$ and one can sharpen \fer{eq:rrdpcvg} as follows (see Proposition \ref{prop:psiform} and equation \fer{eq:mqdecay}): There are constants $C,\alpha>0$, and there is a random variable $n_0(\po)$ with ${\mathbb E}[\e^{\alpha n_0}]<\infty$, s.t. for all $\po\in\Omega_1$ and all $n\geq n_0(\po)$, we have $\| M(\omega_1)\cdots M(\omega_n) - |\psi_\cS\rangle\langle\psi^*_\cS| \ \|\leq C\e^{-\alpha n}$.

While this result allows us to study the large time behaviour of
observables of the small system $\cS$ (see Section
\ref{ssec:systobsreduc}), in order to study the physically relevant instantaneous observables, we need to understand products of the form \fer{mgre6}. In our probabilistic setting, they read $M(\omega_1)\cdots M(\omega_n) N(T^n\po)$.

%%%%%%%%%%%%% PROP: CONVERGENCE M_1...M_n N_{n+1} %%%%%%%%%%%%%%%%%%%%%%%%%%%%%%%%%
\begin{thm}
\label{mgrethm1}
Let $M(\omega)$ be a random reduced dynamics operator 
and let $N(\po)$ be a random matrix, uniformly bounded in $\po$. Suppose that 
$\p(M(\omega)\in\cme)\neq 0$. Then there exists a set $\Omega_2\subset
 \Omega^{\N^*}$ s.t. ${\mathbb P}(\Omega_2)=1$ and s.t. for any $\po\in\Omega_2$,
\begin{equation}
\lim_{n\rightarrow\infty}\frac 1\nu\sum_{n=1}^\nu M(\omega_1)\cdots
M(\omega_n) N(T^n\po)= |\psi_\cS\ket\bra \theta|\ \E[N].
\label{eq:instrrdp}
\end{equation}
\end{thm}
%%%%%%%%%%%%%%%%%%%%%%%%%%%%%%%%%%%%%%%%%%%%%%%%%%%%%%%%%%%%%%%%%%%%%%%%%%%%%%%%%%%

\noindent
{\bf Remark.\ } In our dynamical process, $N(\po)$ depends only on finitely many variables $\omega_{m-l},\ldots,\omega_{m+r}$, see \fer{ennemm}, so measurability and boundedness of the random matrix $N$ are easily established in concrete applications.

\noindent
{\bf Proof of Theorem \ref{mgrethm1}.} Using the decomposition \fer{eq:psi} together with
\fer{eq:mqdecay}, it suffices to show that 
\be
\lim_{\nu\rightarrow\infty}\frac 1\nu\sum_{n=1}^\nu N^*(T^n\po)
\theta_n(\po)= \E[N]^* \theta. \label{mgre5}
\ee
We follow the strategy of \cite{bjm2} used to prove \fer{eq:mqdecay1} of the present paper. From (\ref{eq:theta1}) we get 
\bea
\label{eq:thetaerg} 
\lefteqn{\hspace*{-2.8cm} \sum_{n=1}^\nu N^*(T^n\po)\theta_n(\po)= \sum_{n=1}^\nu \sum_{j=0}^{n-1} N^*(T^n\po)M_Q^*(T^{n-1}\po)
 \cdots M_Q^*(T^{j+1}\po)\psi(T^j\po) }\nonumber\\
\qquad \qquad \ & = & \! \sum_{k=1}^\nu \sum_{j=0}^{\nu-k}
  N^*(T^{k+j}\po)M_Q^*(T^{k+j-1}\po)\cdots M_Q^*(T^{j+1}\po)
  \psi(T^j\po).
\eea 
Let us introduce the random vectors 
\be\label{thek} 
\theta^{(k)}(\po) = N^*(T^k\po) M_Q^*(T^{k-1}\po)M_Q^*(T^{k-2}\po)\cdots
M_Q^*(T\po)\psi(T^0\po), 
\ee 
so that, by (\ref{eq:thetaerg}), 
\bea
\label{eq:thetaerg2}
\frac{1}{\nu}\sum_{n=1}^\nu N^*(T^n\po)\theta_{n}(\po) & = &  \sum_{k=1}^\nu\frac{1}{\nu}\sum_{j=0}^{\nu-k}
  \theta^{(k)}(T^j\po) \nonumber\\
& = & \sum_{k=1}^\infty \chi_{\{k\leq \nu\}}\sum_{j=0}^{\nu-k}\theta^{(k)}(T^j\po)\frac{1}{\nu} =:
\sum_{k=1}^\infty g(k,\nu, \po). 
\eea 
For each fixed $k$, by ergodicity, there exists a set $\Omega_{(k)}\subset\Omega^{\N^*}$
of probability one, such that, for all $\po\in \Omega_{(k)}$, the
following limit exists 
\bea 
\lim_{\nu\ra\infty}g(k,\nu, \po) & = & \lim_{\nu\ra\infty}
  \frac{1}{\nu-k+1}\sum_{j=0}^{\nu-k}\theta^{(k)}(T^j\po)\frac{\nu-k+1}{\nu}\nonumber\\
 & = & \lim_{J\ra\infty}\frac{1}{J+1}\sum_{j=0}^{J}\theta^{(k)}(T^j\po)
  ={\mathbb E}[\theta^{(k)}].\nonumber 
\eea 
Therefore, on the set $\Omega_{\infty}:=\cap_{k\in\mathbb N}\Omega_{(k)}$ of probability
one,  for any $k\in\mathbb N$, we have by independence of the
$M(\omega_j)$, $1\leq j\leq k$, and of $N^*(T^k\po)$,
\be\label{eq:limg} 
\lim_{\nu\ra\infty} g(k,\nu, \po)=\E[\theta^{(k)}]= \E[N^*]\, \E[M_Q^*]^{k-1}\, \E[\psi]. 
\ee
It follows from Proposition \ref{prop:psiform}, Theorem \ref{thm:bjm2} and the boundedness of $N(\po)$ that for $\po\in \Omega_2=\Omega_1 \cap\Omega_{\infty}$, we have $\|\theta^{(k)}(T^j\po)\|\leq C\e^{\alpha n_0(T^j\po)} \e^{-\alpha (k-1)}$. Therefore, for all $\nu$ large enough, and for all $1\leq k\leq\nu$,
\begin{equation}
\|g(k,\nu,\po)\| \leq C\frac{1}{\nu}\sum_{j=0}^{\nu-k} \e^{\alpha n_0(T^j\po)} \e^{-\alpha(k-1)}\leq 2C {\mathbb E}[\e^{\alpha n_0}] \e^{-\alpha(k-1)},
\label{uppbnd}
\end{equation}
where we have used ergodicity in the last estimate. Of course, the same upper bound \fer{uppbnd} holds for $k>\nu$, since then $g(k,\nu,\po)=0$. The r.h.s. of \fer{uppbnd} is summable w.r.t. $k\in{\mathbb N}$, so we can use the Lebesgue Dominated Convergence Theorem in (\ref{eq:thetaerg2}) to
conclude that, almost surely on $\Omega_2$, $\lim_{\nu\to\infty} \frac{1}{\nu}\sum_{n=1}^\nu
N^*(T^n\po)\theta_n(\po)= {\mathbb E}[N^*] \sum_{k=0}^\infty
{\mathbb E}[M^*_Q]^k\  {\mathbb E}[\psi]$.
Relation \fer{mgre5}, and thus the proof of the theorem, now
follow from (\ref{eq:thetalim}). \qed

%{\it Remark.\ } In the last equality, we have used that the
%instantaneous observables satisfy the condition (IDO) of the last
%section. In the more general case, ${\mathbb E}[N^*]$ may depend
%on $k$. This will influence the argument here below in the
%following way: instead of being able to split the contributions of
%each factor on the r.h.s. of \fer{eq:limg} when summing over $k$,
%this sum will now "entangle" the first two factors. The reader can
%now easily generalize the remainder of the proof, and the
%corresponding final result.

%%%%%%%%%%%%%%%%%%%%%%%%%%%%%%%%%%%%%%%%%%%%%%%%%%%%%%%%%%%%%%%%%%%%%%%%%%%%%%%%%
%%%%%%%%%%%%%%%%%%%%%%%%%%%%%%%%%%%%%%%%%%%%%%%%%%%%%%%%%%%%%%%%%%%%%%%%%%%%%%%%%
%%%%%%%%%%%%%%%%%%%% SECTION: ASYMPTOTIC STATE %%%%%%%%%%%%%%%%%%%%%%%%%%%%%%%%%%
%%%%%%%%%%%%%%%%%%%%%%%%%%%%%%%%%%%%%%%%%%%%%%%%%%%%%%%%%%%%%%%%%%%%%%%%%%%%%%%%%
%%%%%%%%%%%%%%%%%%%%%%%%%%%%%%%%%%%%%%%%%%%%%%%%%%%%%%%%%%%%%%%%%%%%%%%%%%%%%%%%%

\section{Proof of Theorem \ref{thmintro2}}
\label{sec:asympstate}

Let $\phi$ be a normalized vector in $\cH$. Fix $\epsilon>0$ and $\po\in\Omega_{\rm ext}$. There exists a $B'=B'(\epsilon,\po)\in\fm'$ of the form \fer{bpn} (with $N$ depending on $\epsilon,\po$), s.t. 
\begin{equation}
\label{mstj2}
\|\phi-B'\psi_0\|<\epsilon.
\end{equation}
Here, both $\phi$ and $\psi_0$ may depend on $\po$. It follows that 
\begin{equation}
\big|\scalprod{\phi}{\alpha_{\po}^m(O)\phi} - \scalprod{B'\psi_0}{\alpha_{\po}^m(O)B'\psi_0}\big| < 2\epsilon \, \|O\|.
\label{mstj1}
\end{equation}
Using  
Proposition \ref{prop:systobsreduc} and Theorem \ref{mgrethm1}, and that $B'$ commutes with $\alpha_{\po}^m(O)$,  we 
arrive at the relations
\begin{eqnarray}
\lefteqn{\lim_{\mu\rightarrow\infty} \frac 1\mu \sum_{m=1}^\mu 
 \scalprod{\psi_0}{(B')^* B' \alpha_{\po}^m(O) \psi_0} = \lim_{\mu\rightarrow\infty} \frac 1\mu \sum_{m=N+1}^\mu
\scalprod{\psi_0}{(B')^* B' \alpha_{\po}^m(O) \psi_0}}\nonumber\\
 &=& \lim_{\mu\to\infty} \frac 1\mu\sum_{m=N+1}^\mu
  \left\langle\psi_0,  (B')^*  B' ({\tilde U}_N^+)^*
   \e^{\ri \tau(\omega_1){K(\omega_1)}}\cdots
\e^{\ri\tau(\omega_N){K(\omega_N)}}P\times \right. \nonumber\\
& & \qquad \times  M(\omega_{N+1})
 \cdots M(\omega_{m-l-1}) N(\omega_{m-l},\omega_{m-l+1},\ldots,\omega_{m+r})
    \psi_0\Big\ket\nonumber\\
&=& \scalprod{\psi_0}{(B')^*B'\psi_0}\ \scalprod{\theta}{{\mathbb
E}[N(O)]\psi_\cS},\label{mgre21}
 \end{eqnarray}
for all $\po$ in a set $\Omega_2$ of measure one. It follows from \fer{mstj2} that $(1-\epsilon)^2 < \scalprod{\psi_0}{(B')^*B'\psi_0}= \|B'\psi_0\|^2 < 
 (1+\epsilon)^2$. Since $\epsilon$ is arbitrary, using the latter bound in \fer{mgre21} and taking into account \fer{mstj1}, we conclude that \fer{intro-1'} holds for any vector initial state $\varrho(\cdot)=\scalprod{\phi}{\cdot \,\phi}$. Finally, the argument leading to \fer{eq:asympstate} shows that \fer{intro-1'} holds for all normal initial states. The proof of Theorem \ref{thmintro2} is complete.
\qed

%%%%%%%%%%%%%%%%%%%%%%%%%%%%%%%%%%%%%%%%%%%%%%%%%%%%%%%%%%%%%%%%%%%%%%%%%%%%%%%%%
%%%%%%%%%%%%%%%%%%%%%%%%%%%%%%%%%%%%%%%%%%%%%%%%%%%%%%%%%%%%%%%%%%%%%%%%%%%%%%%%%
%%%%%%%%%%%%% SECTION: ENERGY AND ENTROPY %%%%%%%%%%%%%%%%%%%%%%%%%%%%%%%%%%%%%%%
%%%%%%%%%%%%%%%%%%%%%%%%%%%%%%%%%%%%%%%%%%%%%%%%%%%%%%%%%%%%%%%%%%%%%%%%%%%%%%%%%
%%%%%%%%%%%%%%%%%%%%%%%%%%%%%%%%%%%%%%%%%%%%%%%%%%%%%%%%%%%%%%%%%%%%%%%%%%%%%%%%%

\section{Proof of Theorem \ref{thmintro4}}
\label{sec:energy-entropy}

An easy application of Theorem \ref{thmintro2} shows that for any normal initial state $\varrho$,
\be\label{asymptenergy}
\lim_{m\to\infty} \frac{\varrho (\Delta E(m,\po))}{m}= \varrho_+(j_+),\ \ a.s.,
\ee
where
\be\label{asymptenergy2}
j_+=\E \left[ PVP- P\e^{\ri \tau L}V\e^{-\ri \tau L} P \right]=\E
\left[ P(V-\alpha^\tau(V))P\right].
\ee
The energy grows linearly in time almost surely, at the rate $\d E_+$.\footnote{The definition of $\d E_+$ differs from the one of \cite{bjm} by a factor $\frac{1}{\tau}$: here $\d E_+$ represent the asymptotic average energy production per interaction and not per unit of time. One could also study the average energy production per unit of time. It is easy to see that 
$$\lim_{m\to\infty} \frac{\varrho (\Delta E(m,\po))}{\tau(\omega_1)+\cdots +\tau(\omega_m)}=\frac{\d E_+}{\E[\tau]}, \quad a.s.
$$}
In order to show the expression for $\d E_+$ given in Theorem \ref{thmintro4}, it suffices to prove that $\varrho_+[\E(P(L_\cS-\alpha^\tau(L_\cS))P)]=0$.

Let $\varrho$ be a normal state. Although $L_\cS\notin \fm$, still $\alpha^k(L_\cS)-\alpha^{k-1}(L_\cS)$ is an instantaneous observable belonging to $\fm$. This follows from $\e^{\ri \tau_kL_k}L_\cS \e^{-\ri \tau_kL_k}-L_\cS\in \fm_\cS\otimes \fm_{\cE_k}$, which in turn is proven by noting that 
$$\e^{\ri \tau_kL_k}L_\cS \e^{-\ri \tau_kL_k}-L_\cS = \int_0^{\tau_k} \e^{\ri t L_k}[\ri L_k,L_\cS] \e^{-\ri t L_k} \d t
 = \int_0^{\tau_k} \e^{\ri t L_k}[\ri V_k,L_\cS] \e^{-\ri t L_k} \d t,
$$
where $[\ri V_k,L_\cS]=-\frac{\d}{\d t} \e^{\ri t L_\cS}V_k\e^{-\ri t L_\cS}|_{t=0}\in \fm_\cS\otimes \fm_{\cE_k}$. 

As a consequence, $\alpha^k(L_\cS)-\alpha^{k'}(L_\cS)\in\fm$, and we can apply Theorem \ref{thmintro2} to obtain $\lim_{m\to\infty}\frac{1}{m} \sum_{k=1}^m \varrho \left( \alpha^k(L_\cS)-\alpha^{k+1}(L_\cS) \right)=\varrho_+\big(\E[P(L_\cS-\alpha^\tau(L_\cS))P]\big)\ \ a.s.$\  
On the other hand, we have that $\sum_{k=1}^m \varrho \left( \alpha^k(L_\cS)-\alpha^{k+1}(L_\cS) \right)=\frac{1}{m}\varrho\left( \alpha^1(L_\cS)-\alpha^{m+1}(L_\cS) \right)$, 
which tends to zero as $m\rightarrow\infty$. This proves the formula for $\d E_+$ given in Theorem \ref{thmintro2}.

Next we show the expression for $\d S_+$ in Theorem \ref{thmintro2}. The following result is deterministic, we consider $\po$ fixed and do not display it.
\begin{prop}\label{prop:energyentropy} Let $\varrho$ be a normal state
  on $\fm$. Then for any $m\geq 1$, we have
\bea
\lefteqn{\Delta S(m)  :=   {\rm Ent} (\varrho\circ\alpha^m| \varrho_0)-{\rm Ent} (\varrho|
\varrho_0)}\nonumber\\
 & \! = \! & \! \varrho \left( \sum_{k=1}^m \beta_{\cE_k}
  \left(j(k)+ \alpha^{k-1}(L_\cS+V_k)-\alpha^k(L_\cS+V_{k+1})\right) +\beta_\cS (
  \alpha^m(L_\cS)-L_\cS) \right),\nonumber
\eea
where the energy jump $j(k)$ has been defined in \fer{def:energy}. 
\end{prop}

\proof The proof is similar to that one of Proposition 2.6. in
\cite{bjm}. Using the entropy production formula \cite{JP1}, we have
\be\label{eq:entropyprod}
\Delta S(m) = \varrho \left(\alpha^m \left(\beta_\cS L_\cS+\sum_k
    \beta_{\cE_k}L_{\cE_k} \right) - \sum_k
    \beta_{\cE_k}L_{\cE_k}-\beta_\cS L_\cS \right).
\ee
Clearly, the sums in the argument of $\varrho$ in the right hand side only extend from $k=1$ to
$k=m$. We examine the difference of the two terms with index $k$.
\bea
\alpha^m(\betak L_{\cE_k})-\betak L_{\cE_k} & = & \alpha^k(\betak L_{\cE_k})-\betak L_{\cE_k}\nonumber\\
 & = & \betak \alpha^k(L_k)-\betak L_{\cE_k}-\betak
 \alpha^k(L_\cS+V_k)\nonumber\\
 & = & \betak \alpha^{k-1}(L_k)-\betak L_{\cE_k}+\betak j(k)-\betak
 \alpha^k(L_\cS+V_{k+1})\nonumber\\
 & = & \betak \alpha^{k-1}(L_\cS+V_k)+\betak j(k)-\betak
 \alpha^k(L_\cS+V_{k+1}),\nonumber
\eea
where we use $\alpha^m( L_{\cE_k})=\alpha^k( L_{\cE_k})$ in the first step, $\alpha^k(L_k)=\alpha^{k-1}(L_k)$ and \fer{def:energy} in the third step, and in the last one $\alpha^{k-1}(L_{\cE_k})=L_{\cE_k}.$
\qed

By Proposition \ref{prop:energyentropy}, we have for all $m\geq 1$, $\po\in\Omega_{\rm ext}$
\begin{eqnarray*}
\frac{\Delta S(m,\po)}{m} & = & \frac{1}{m}\sum_{k=1}^m \varrho \left(\beta_{\cE_k}
  j(k)\right)+ \frac{1}{m}\sum_{k=1}^m \varrho \left(\beta_{\cE_k} \left(\alpha^{k-1}(V_k)-\alpha^k(V_{k+1})\right) \right) \\
 & & + \frac{1}{m}\sum_{k=1}^m \varrho \left(\beta_{\cE_k} \left( \alpha^{k-1}(L_\cS)-\alpha^k(L_\cS)\right) \right)+\frac{1}{m}\varrho (\beta_\cS ( \alpha^m(L_\cS)-L_\cS)  ). 
\end{eqnarray*}
Using Theorem \ref{thmintro2} we see that with probability one (and where $M$ denotes the reduced dynamics operator)
\begin{eqnarray*}
\lim_{m\to\infty} \frac{\Delta S(m,\po)}{m} & = & \varrho_+\Big( \E[\beta_\cE M]\,
\E[PVP]-\E[\beta_\cE P\alpha^\tau(V) P]
\Big)+\varrho_+\Big(\E[\beta_\cE PVP] \\
 & &  -\E[\beta_\cE M]\, \E[PVP]\Big) +\varrho_+\Big(\E[\beta_\cE P(L_\cS-\alpha^\tau(L_\cS))P]\Big) \\
 & = & \varrho_+ \Big(\E\big[\beta_\cE P((L_\cS+V)-\alpha^\tau(L_\cS+V))P\big] \Big).
\end{eqnarray*}
This completes the proof of Theorem \ref{thmintro4}. \qed

%%%%%%%%%%%%%%%%%%%%%%%%%%%%%%%%%%%%%%%%%%%%%%%%%%%%%%%%%%%%%%%%%%%%%%%%%%%%%%%%%%%%%%%%%%%%%
%%%%%%%%%%%%%%%%%%%%%%%%%%%%%%%%%%%%%%%%%%%%%%%%%%%%%%%%%%%%%%%%%%%%%%%%%%%%%%%%%%%%%%%%%%%%%
%%%%%%%%%%%%%%%%%%%%%%%%%%%%%%%%%%%%%%%%%%%%%%%%%%%%%%%%%%%%%%%%%%%%%%%%%%%%%%%%%%%%%%%%%%%%%
%%%%%%%%%%%%%%%%%%%%%%%%%%%%%%%%%%%%%%%%%%%%%%%%%%%%%%%%%%%%%%%%%%%%%%%%%%%%%%%%%%%%%%%%%%%%%

\section{Spin-spin models and proof of Theorem \ref{thmintro5}} 
\label{sec:spin}

In this section, we consider both $\cS$ and $\cE$ to be two-level systems, with interaction given by \fer{eqnv}.
This is a particular case of the third example in \cite{bjm}. The
main results of this section have been anounced in \cite{bjm2}.

The observable algebra for $\cS$ and for $\cE$ is $\fa_\cS=\fa_\cE=M_2(\C)$. 
Let $E_\cS, E_\cE>0$ be the ``excited'' energy level of $\cS$ and of $\cE$, respectively. Accordingly, the Hamiltonians are given by
$$
h_\cS= \left[ \begin{array}{cc} 0 & 0 \\ 
0 & E_\cS \end{array} \right] 
\mbox{\ \ and \ \ }
h_\cE= \left[ \begin{array}{cc} 0 & 0 \\ 
0 & E_\cE \end{array} \right].
$$ 
The dynamics are given by $\alpha_\cS^t(A)= \e^{i t h_\cS}A\e^{-i t h_\cS}$  and  $\alpha_\cE^t(A)= \e^{i t h_\cE}A\e^{-i t h_\cE}$. 
We choose (for computational convenience) the reference state of $\cE$ to be the Gibbs state at inverse temperature $\beta$, see \fer{mm110}, and we choose the reference state for $\cS$ to be the tracial state, $\varrho_{0,\cS}(A)=\frac{1}{2}\tr(A)$. The interaction operator is defined by $\lambda v$, where $\lambda$ is a coupling constant, and $v$ is given in \fer{eqnv}. The creation and annihilation operators are represented by the matrices 
$$
a_{\#}= \left[ \begin{array}{cc} 0 & 1 \\ 
0 & 0 \end{array} \right]
\mbox{\ \ and\ \ }
a_{\#}^*= \left[ \begin{array}{cc} 0 & 0 \\ 
1 & 0 \end{array} \right].
$$
The Heisenberg dynamics of $\cS$ coupled to one element $\cE$ is given by the $*$-automorphism group $t\mapsto \e^{ith_\lambda} Ae^{-ith_\lambda}$, $A\in{\frak A}_\cS\otimes{\frak A}_\cE$, $h_\lambda = h_\cS+h_\cE+\lambda v$.

To find a Hilbert space description of the system, one performs the GNS construction of $(\fa_\cS,\varrho_{0,\cS})$ and $(\fa_\cE,\varrho_{\beta,\cE})$, see e.g. \cite{BR,bjm}. In this representation, the Hilbert spaces are given by  
$\cH_\cS = \cH_\cE = {\mathbb C}^2\otimes{\mathbb C}^2$, the Von
Neumann algebra by $\fm_\cS=\fm_\cE=M_2(\C)\otimes \one_{\C^2}\subset \cB(\C^2\otimes\C^2),$
and the vectors representing $\varrho_{0,\cS}$ and $\varrho_{\beta,\cE}$ are $\psi_\cS = \frac{1}{\sqrt 2} \left(|0\ket\otimes|0\ket+|1\ket\otimes|1\ket\right)$ and $\psi_\cE=\frac{1}{\sqrt{{\rm Tr}\e^{-\beta h_\cE}}}\left( |0\ket\otimes|0\ket + \e^{-\beta E_\cE/2}|1\ket\otimes|1\ket\right)$, respectively, i.e., we have
$\varrho_{\beta_{\#},\#}(A)=\scalprod{\psi_{\#}}{(A\otimes\bbbone)\psi_{\#}}$,
$\#=\cS,\cE$, $\beta_\cE=\beta$, $\beta_\cS=0$, and where
$|0\ket$ (resp. $|1\ket$) denote the ground (resp. excited) state of
$h_\cS$ and $h_\cE$.  
Finally, the Liouvillean $L$ is given by  
\begin{eqnarray*}
L & = & (h_\cS\otimes \one_{\C^2}-\one_{\C^2}\otimes h_\cS)\otimes 
 (\one_{\C^2}\otimes\one_{\C^2})+(\one_{\C^2}\otimes\one_{\C^2})\otimes
 (h_\cE\otimes \one_{\C^2}-\one_{\C^2}\otimes h_\cE)\nonumber\\
 & & + \lambda(a_\cS\otimes\one_{\C^2})\otimes(a^*_\cE\otimes
 \one_{\C^2})
 +\lambda(a_\cS^*\otimes\one_{\C^2})\otimes(a_\cE\otimes \one_{\C^2}).
\end{eqnarray*}

%%%%%%%%%%%%%%%%%%%%%%%%%%%%%%%%%%%%%%%%%%%%%%%%%%%%%%%%%%%%%%%%%%%%%%%%%%%%%%%%%%%%%%%%%%%%%
%%%%%%%%%%%%%%%%%%%%%%%%%%%%%%%%%%%%%%%%%%%%%%%%%%%%%%%%%%%%%%%%%%%%%%%%%%%%%%%%%%%%%%%%%%%%%

\subsection{Spectral analysis of the reduced dynamics operator $M$}\label{ssec:spectrum}

The RDO $M$ is defined by \fer{defmj}. However, in this example, where
the hamiltonian $h_\lambda$ is explicitly diagonalizable, we shall use
another expression for
it, which may look less simple but has the advantage that it only
makes use of the self-adjoint hamiltonian. Since $\psi_\cS$ is cyclic for $\fm_\cS$ and $\cH_\cS$
has finite dimension, $\forall \phi\in\cH_\cS, \ \exists!
A_\cS=A\otimes \one_{\C^2}\in\fm_\cS$ such that
$\phi=A_\cS\psi_\cS$. It is then easy to see that 
\be
M (A\otimes\one_{\C^2}) \psi_\cS = (\cM(A)\otimes\one_{\C^2}) \psi_\cS,
\ee
and where the map $\cM$ acts on $\fa_\cS$ and is defined as
%%%%%%%%%%%%%%%%%%%%%%%% DEF: \cal{M} MAP %%%%%%%%%%%%%%%%%%%%%%%%%%%%%%%%%%%%%%%%%%
\be\label{def:cm}
\cM(A):=\tr_\cE \left( e^{i\tau h_\lambda}\, A\otimes\one \, e^{-i\tau h_\lambda} \right),
\ee
%%%%%%%%%%%%%%%%%%%%%%%%%%%%%%%%%%%%%%%%%%%%%%%%%%%%%%%%%%%%%%%%%%%%%%%%%%%%%%%%%%
where $\tr_\cE(A_\cS\otimes A_\cE):=\varrho_{\beta,\cE}(A_\cE)A_\cS$
denotes the partial trace over $\cE$.

Similarly, if $\cM^*$ denotes the map dual to $\cM$, i.e. $\forall \rho,A\in M_2(\C)$, $\tr(\rho \cM(A))=\tr(\cM^*(\rho)A),$ then we have, for any density matrix $\rho$
\be\label{linkmcalm}
((\cM^*(\rho))^*\otimes \one) \psi_\cS=M^* (\rho^*\otimes \one)\psi_\cS.
\ee
In particular, the spectrum of the map $\cM^*$ is in one-to-one correspondance with the spectrum of the operator $M^*$ (via complex conjugation), and if $\rho$ is an eigenvector of $\cM^*$ for the eigenvalue $1$ (which we know to exist), then the ``corresponding eigenvector'' of $M^*$ is $\psi_\cS^*=(\rho^*\otimes \one) \psi_\cS$. A simple computation shows that the four eigenvalues of $h_\lambda$ are $E_{0+}=0$, $E_{0-}=E_\cS+E_\cE$ and
\be\label{def:hlambdaev}
E_{1\pm}=\frac{1}{2}(E_\cS+E_\cE) \pm \frac{1}{2}\sqrt{(E_\cS-E_\cE)^2+4\lambda^2}.
\ee
The corresponding normalized eigenvectors are given by $\psi_{0+}=|0\ket\otimes |0\ket$, $\psi_{0-}=|1\ket\otimes |1\ket$, and $\psi_{1\pm}= a_{1\pm} |1\ket\otimes |0\ket + b_{1\pm} |0\ket\otimes |1\ket$, respectively, where 
\be\label{abrelation}
a_{1\pm}=-\frac{\lambda}{\sqrt{\lambda^2+(E_\cS-E_{1\pm})^2}}, \ \ b_{1\pm}=\frac{E_\cS-E_{1\pm}}{\sqrt{\lambda^2+(E_\cS-E_{1\pm})^2}}.
\ee
We finally denote $a_{0+}=b_{0-}=1$ and $a_{0-}=b_{0+}=0$. Inserting the spectral decomposition of $h_\lambda$ into (\ref{def:cm}) gives the following result.
%%%%%%%%%%%%%%%%%%%% LEMMA: FORMULAE FOR \cM and \cM^* %%%%%%%%%%%%%%%%%%%%%%%%%%%%%
\begin{lem}\label{lem:spinmformula} For any $A\in\fa$,
\bea\label{eq:spinm}
\cM(A) & = & Z_{\beta,\cE}^{-1}\sum_{n,\sigma,n',\sigma'} e^{i\tau(E_{n\sigma}-E_{n'\sigma'})} \left( \bar{a}_{n\sigma}a_{n'\sigma'} \bra n|An'\ket+  \bar{b}_{n\sigma}b_{n'\sigma'} \bra 1-n|A(1-n')\ket\right) \nonumber \\
 & & \qquad \qquad \quad \times \left(  a_{n\sigma}\bar{a}_{n'\sigma'} |n\ket\bra n'|+ e^{-\beta E_\cE} b_{n\sigma}\bar{b}_{n'\sigma'} | 1-n \ket\bra 1-n'|\right),
\eea
where $n,n'\in\{0,1\}$ and $\sigma,\sigma'\in\{-,+\}$ and $Z_{\beta,\cE}=\tr(e^{-\beta h_\cE}).$ Similarly, for any density matrix $\rho$,
\bea\label{eq:spinmstar}
\cM^*(\rho) \! \! & = & \!\! Z_{\beta,\cE}^{-1}\sum_{n,\sigma,n',\sigma'} e^{i\tau(E_{n\sigma}-E_{n'\sigma'})} \left( a_{n\sigma}\bar{a}_{n'\sigma'} \bra n'|\rho n\ket+  e^{-\beta E_\cE} b_{n\sigma}\bar{b}_{n'\sigma'} \bra 1-n'|\rho(1-n)\ket\right) \nonumber \\
 & & \qquad \qquad \quad \times \left(  \bar{a}_{n\sigma}a_{n'\sigma'} |n'\ket\bra n|+ \bar{b}_{n\sigma}b_{n'\sigma'}
 | 1-n' \ket\bra 1-n|\right).
\eea
\end{lem}
%%%%%%%%%%%%%%%%%%%%%%%%%%%%%%%%%%%%%%%%%%%%%%%%%%%%%%%%%%%%%%%%%%%%%%%%%%%%%%%%%%

The above lemma allows us to make a complete spectral analysis of
$M$. 
%%%%%%%%%%%%%%%%%%%%%%%% PROP: SPECTRAL DECOMPOSITION OF M %%%%%%%%%%%%%%%%%%%%%%%
\begin{prop}\label{prop:spectm} \ \
\begin{itemize}
\item[1.] The eigenvalues of $M$ are $1, e_0,e_-,e_+$ where $e_0$ is given in \fer{enot}, 
\begin{eqnarray*}
e_- & = & \frac{\left(E_\cS-E_\cE-\sqrt{(E_\cS-E_\cE)^2+4\lambda^2}\right)^2+4\lambda^2 e^{\ri\tau \sqrt{(E_\cS-E_\cE)^2+4\lambda^2}}}{\left(E_\cS-E_\cE-\sqrt{(E_\cS-E_\cE)^2+4\lambda^2}\right)^2+4\lambda^2}\\
 & & \qquad\qquad\qquad\qquad\qquad\qquad\qquad\qquad  \times\e^{\ri\tau (E_\cS+E_\cE-\sqrt{(E_\cS-E_\cE)^2+4\lambda^2})},\\
e_+ & = & \overline{e_-}.
\end{eqnarray*}
Moreover, the eigenstates of $M^*$ for the eigenvalues $1, e_0,e_-,e_+$ are respectively
$\psi_\cS^*=(\e^{-\tbeta h_\cS}\otimes \one_{\C^2})\psi_\cS$ where $\tbeta:=\beta E_\cE/E_\cS$, $\phi_0=|0\ket\otimes |0\ket - |1\ket\otimes |1\ket,$ $\phi_-=|0\ket\otimes |1\ket$ and $\phi_+=|1\ket\otimes |0\ket$.

\item[2.] The functions $|e_0(\tau)|,$ $|e_+(\tau)|$ and $|e_-(\tau)|$ are continuous and periodic of period $T:=\frac{2\pi}{\sqrt{(E_\cS-E_\cE)^2+4\lambda^2}}$. Moreover, they have modulus strictly less than $1$ if and only if $\tau\notin T\N.$
\end{itemize}
\end{prop}
%%%%%%%%%%%%%%%%%%%%%%%%%%%%%%%%%%%%%%%%%%%%%%%%%%%%%%%%%%%%%%%%%%%%%%%%%%%%%%%%%%%

\begin{rem}
Since $e_0$ is positive, point 2. proves that $1$ is a non degenerate eigenvalue for $\cM^*$ if and only if $\tau\notin T\N,$ i.e. for all but a discrete set of interaction times.
This condition agrees with the corresponding assumption of \cite{bjm} in the perturbative regime.
\end{rem}

\proof Point 2. follows from point 1. Point 1. is proven by direct computation using (\ref{def:hlambdaev})-(\ref{abrelation})-(\ref{eq:spinmstar}). \qed

%%%%%%%%%%%%%%%%%%%%%%%%%%%%%%%%%%%%%%%%%%%%%%%%%%%%%%%%%%%%%%%%%%%%%%%%%%%%%%%%%%%%%%%%%%%
%%%%%%%%%%%%%%%%%%%%%%%%%%%%%%%%%%%%%%%%%%%%%%%%%%%%%%%%%%%%%%%%%%%%%%%%%%%%%%%%%%%%%%%%%%%

\subsection{Proof of Theorem \ref{thmintro5}}
\label{ssec:spincvg}

Point 2. of Proposition \ref{prop:spectm} shows that $M\in\cme$ if and only if $\tau\notin T\N$. Hence, for this spin-spin model, Theorem \ref{thmintro2} applies if and only if $\p(\tau\notin T\N)\neq 0$, which is precisely the assumption we have in each of the three situations of Theorem \ref{thmintro5}. It remains to compute the asymptotic state $\varrho_+$ in each of these three situation. Using the complete spectral decomposition of $M^*(\omega)$ (see Proposition \ref{prop:spectm}), we compute explicitly its expectation $\E[M^*]$ and then the spectral projection $P_{1,\E[M^*]}$. After computation, we get:
\begin{itemize}
 \item[1.] {\rm Random interaction time:} $P_{1,\E[M^*]}=\frac{2}{\tr(\e^{-\tbeta h_\cS})}|\psi_\cS\ket\bra \e^{-\tbeta h_\cS}\otimes\one_{\C^2} \ \psi_\cS|$,
 \item[2.] {\rm Random excitation energy of $\cE$:} $P_{1,\E[M^*]}=|\psi_\cS \ket\bra \rho_E \otimes \one_{\C^2} \ \psi_\cS |,$ where 
\bea
\rho_E & = & \left[1-(1-\E(e_0))^{-1}\E((1-e_0)(1-2
  Z_{\tbeta,\cS}^{-1}))\right] |0\rangle\langle 0| \nonumber\\
 & & \qquad\qquad + \left[1+(1-\E(e_0))^{-1}\E((1-e_0)(1-2
   Z_{\tbeta,\cS}^{-1}))\right] |1\rangle\langle 1|, \nonumber
\eea
 \item[3.] {\rm Random temperature of $\cE$:} $P_{1,\E[M^*]}=\frac{2}{\tr(\e^{-\tilde{\beta} h_\cS})}|\psi_\cS\ket\bra \e^{-\tilde{\beta} h_\cS}\otimes\one_{\C^2} \ \psi_\cS|$, where ${\beta}=-E_\cS^{-1}\log (\E[Z_{\beta'(\omega),\cS}^{-1}]^{-1}-1)$.
\end{itemize}
Combining these formulas with \fer{intro3'} give the various expressions for the asymptotic state $\varrho_+$. Finally, when the interaction time $\tau$ is random, the map $M^*(\omega)$ has a deterministic eigenvector for the eigenvalue $1$. This allows for stronger convergence results as mentioned in the Remark after Theorem \ref{thm:bjm2}.

\section{Spin-Fermion models and proof of Theorem \ref{thmintro3}}
\label{sec:spinfermion}

We combine our convergence results with a rigorous perturbation
theory in the coupling strength between $\cS$ and $\cC$. We take $\cS$ to be a 2-level atom and the $\cE$ are large quantum systems, each one
modeled by an infinitely extended gas of free thermal fermions. The random parameters are the temperature of the system $\cE_k$, $T_k=\beta_k^{-1}$, as well as the interaction time $\tau_k$.

The state space and the reference vector of $\cS$ are
\begin{equation}
\cH_\cS = {\mathbb C}^2\otimes{\mathbb C}^2, \ \ \ \ \psi_\cS =
\frac{1}{\sqrt 2}\left( |0\ket\otimes |0\ket +
|1\ket\otimes |1\ket \right), \label{mgre30}
\end{equation}
where $\{ |0\ket=[1,0]^T, |1\ket=[0,1]^T\}$ is the canonical
basis of ${\mathbb C}^2$. \fer{mgre30} gives the GNS representation of the
trace state on the algebra of complex matrices ${\mathbb
M}_2({\mathbb C})$, ${\textstyle \frac 12} {\rm Tr}(A_\cS) = \scalprod{ \psi_\cS }{
(A_\cS\otimes\bbbone_\cS) \psi_\cS}$, 
for all $A_\cs\in {\mathbb M}_2({\mathbb C})$. The von Neumann
algebra of observables represented on $\cH_\cS$ is thus $\fm_\cS = {\mathbb M}_2(\C)\otimes\one \subset \cB(\C^2\otimes \C^2)$. The Heisenberg dynamics of $\cS$ is given by $\e^{\ri t \sigma_z}
A_\cs \e^{-\ri t\sigma_z}$. The Pauli matrices $\sigma_z$ and
$\sigma_x$ (the latter plays a role in the interaction) are
\begin{equation}
\sigma_z = \left[ \begin{array}{cc}
1 & 0 \\ 0 & -1 \end{array} \right], 
\ \ \ \sigma_x = \left[ \begin{array}{cc}
0 & 1 \\ 1 & 0 \end{array} \right]. \label{paulimat}
\end{equation}
On the algebra $\fm_\cs$,
the dynamics is implemented as $\tau_\cS^t (A_\cS\otimes\one) = \e^{\ri t L_\cS} (A_\cS\otimes\one)\e^{-\ri t L_\cS}$, 
with standard Liouville operator 
\begin{equation}\label{mgre32} 
L_\cS = \sigma_z\otimes\one -\one\otimes\sigma_z.
\end{equation}
Note that $L_\cS\psi_\cS=0$, as required in \fer{m1}. It is easily
verified that the modular operator $\Delta_{\rm S}$ and the modular conjugation $J_{\rm S}$ are given by
\begin{equation}\label{mgre33} 
\Delta_\cS = \one\otimes\one,\ \ \
J_\cS(\psi\otimes\chi) = \overline{\chi}\otimes\overline{\psi},
\end{equation}
for vectors $\psi,\chi\in \C^2$, and where the bar means
taking complex conjugation of coordinates in the canonical basis.

\medskip

We now describe a single element $\cE$ of the chain, a free
Fermi gas at inverse temperature $\beta$ in the thermodynamic
limit. We refer the reader to \cite{BR} for a detailed
presentation. Let $\fh$ and $h$ be the Hilbert space and the
Hamiltonian for a single fermion, respectively. We represent $\fh$
as $\fh=L^2(\R^+,\d\mu(r);{\frak g})$, 
where ${\frak g}$ is an auxiliary Hilbert space, and we take $h$
to be the operator of multiplication by $r\in\R^+$. (See also footnote \ref{fn1} at the end of Section \ref{sec:intro}). The fermionic annihilation and creation operators $a(f)$ and $a^*(f)$ act on the fermionic Fock space
$\Gamma_-(\fh)$. They satisfy the canonical anti-commutation
relations (CAR). As a consequence of the CAR, the operators $a(f)$
and $a^*(f)$ are bounded and satisfy $\|a^\#(f)\|=\|f\|$ where
$a^\#$ stands for either $a$ or $a^*$. The algebra of
observables of a free Fermi gas is the $C^*$-algebra of operators
$\mathfrak A$ generated by $\{a^\#(f)|f\in\fh\}$. The dynamics is
 given by $\tau_{\rm f}^t(a^\#(f))=a^\#(\e^{\i th}f)$, 
where $h$ is the Hamiltonian of a single particle, acting on
$\fh$. It is well known (see e.g. \cite{BR}) that for any
$\beta>0$, there is a unique $(\tau_\f,\beta)-$KMS state
$\varrho_{\beta}$ on $\mathfrak A$, determined by the
two point function $\varrho_{\beta}(a^*(f)a(f))=\langle f,
(1+\e^{\beta h})^{-1}f\rangle. $ Let us denote by $\Omega_\f$
the Fock vacuum vector, and by $N$ the number operator of
$\Gamma_-(\fh)$. We fix a complex conjugation (anti-unitary
involution) $f\to\bar{f}$ on $\fh$ which commutes with the energy
operator $h$. It naturally extends to a complex conjugation on the
Fock space $\Gamma_-(\fh)$ and we denote it by the same symbol,
i.e. $\Phi\to\bar{\Phi}.$

The GNS representation of the algebra $\mathfrak A$ associated to
the KMS-state $\varrho_{\beta}$ is the triple
$(\h_\cE,\pi_{\beta},\psi_\cE)$ \cite{AW} where 
\be
\h_\cE=\Gamma_-(\fh)\otimes\Gamma_-(\fh), \quad \psi_\cE=\Omega_\f\otimes\Omega_\f, 
\ee 
and 
\be
\begin{array}{l}
\pi_{\beta}(a(f)) = a\left(\frac{\e^{\beta h/2}}{\sqrt{1+\e^{\beta
 h}}}f \right)\otimes \one+(-1)^N\otimes a^*\left(\frac{1}{\sqrt{1+
 \e^{\beta h}}}\bar{f} \right)=:a_{\beta}(f),\\
\pi_{\beta}(a^*(f)) = a^*\left(\frac{\e^{\beta
h/2}}{\sqrt{1+\e^{\beta h}}}f \right)\otimes
\one+(-1)^N\otimes a\left(\frac{1}{\sqrt{1+\e^{\beta
h}}}\bar{f} \right)=:a^*_{\beta}(f).
\end{array}
\label{mgre39}
\ee
The von Neumann algebra of observables for an element $\cE$ of the
chain is $\fm_\cE=\pi_{\beta}({\mathfrak A})''$, 
acting on the Hilbert space $\cH_{\cE}$.
The dynamics on $\pi_{\beta}({\mathfrak A})$ is given by
$\tau_\cE^t(\pi_{\beta}(A))=\pi_{\beta}(\tau_\f^t(A))$, it extends
to $\fm_\cE$ in a unique way. The standard Liouville operator is
given by
\be\label{chainliouv1}
L_\cE=\d\Gamma(h)\otimes\one-\one\otimes\d\Gamma(h). 
\ee
Note that $L_\cE\psi_\cE=0$. Finally,
the modular conjugation and the modular operator associated to
$(\fm_\cE,\psi_\cE)$ are 
\be\label{chainmod1} 
J_\cE(\Phi\otimes\Psi)=(-1)^{N(N-1)/2}\bar{\Psi}\otimes
(-1)^{N(N-1)/2}\bar{\Phi}, \quad \Delta_\cE=\e^{-\beta L_\cE}.
\ee

The combined, uncoupled system has product structure, with Hilbert
space $\ch_\cS\otimes\cH_\cE$, algebra $\fm_\cS\otimes\fm_\cE$,
reference state $\psi_\cS\otimes\psi_\cE$. The uncoupled dynamics
is generated by the Liouville operator
\begin{equation}
\label{mgre42} L_0 = L_\cS + L_\cE.
\end{equation}
We now specify the interaction between the small system and the
elements of the chain. Let $g\in\fh$ be a form factor. The
interaction operator is given by 
\be\label{interaction2}
V:=\sigma_x\otimes\one_{\C^2}\otimes(a_{\beta}(g)+a^*_{\beta}(g))
\ee 
(where $\sigma_x$ is defined in \fer{paulimat}). It produces energy
exchange processes between $\cS$ and $\cE$. Using \fer{mgre33},
\fer{chainmod1}, one readily calculates
\begin{eqnarray}
\lefteqn{ (\Delta_\cS \otimes\Delta_\cE)^{1/2} V  (\Delta_\cS
\otimes\Delta_\cE)^{-1/2} }\nonumber\\
& = & \sigma_x\otimes\one_{\C^2}\otimes\left[
a^*\left(\frac{1}{\sqrt{1+\e^{\beta h}}}g\right)
\otimes\one+a\left(\frac{\e^{\beta h}}{\sqrt{
1+\e^{\beta h}}}g\right)\otimes \one \right.\nonumber\\
 & & \quad \left. +(-1)^N\otimes a^*\left(\frac{\e^{\beta h/2}}{
 \sqrt{1+\e^{\beta h}}}\bar{g}\right)+(-1)^N\otimes
 a\left(\frac{\e^{-\beta h/2}}{\sqrt{1+\e^{\beta
  h}}}\bar{g}\right) \right].
\label{mgre35}
\end{eqnarray}
We assume that $\e^{\beta h/2}g\in{\mathfrak h}$. Then
\fer{mgre35} shows that $(\Delta_\cS\otimes\Delta_\cE)^{1/2} V
(\Delta_\cS\otimes\Delta_\cE)^{-1/2}\in\fm_\cS\otimes\fm_\cE$,
i.e., Condition (A2) of Section \ref{sec:reduction} is satisfied.

%%%%%%%%%%%%%%%%%%%%%% THM: SPIN-FERMION %%%%%%%%%%%%%%%%%%%%%%%%%%%%%%%%%%%%%%%%%%
\begin{thm}[Convergence to asymptotic state]
\label{mgrethm2}
Let $0<\tau_{\rm min}<\tau_{\rm max}<\infty$ and $0<\beta_{\rm max}<\infty$ be given. Let $\tau:\Omega\mapsto[\tau_{\rm min},\tau_{\rm max}]$ and $\beta:\Omega\mapsto(0,\beta_{\rm max}]$ be  random variables. Suppose that $\|(1+\e^{\beta_{\rm max} h/2})g\|<\infty$, and that there is a $\delta>0$ such that 
\begin{equation}
\label{mgre36} 
{\rm p}\big( 
{\rm dist}(\tau,{\textstyle \frac{\pi}{2}{\mathbb N}})>\delta
\big) \neq 0.
\end{equation}
Then there is a constant $\lambda_0>0$, depending on $\tau_{\rm
 min}$, $\tau_{\rm max}$, $\beta_{\rm max}$, $\delta$, and on the form factor $g$, s.t.
 if $0<|\lambda|<\lambda_0$, then ${\rm p}(M(\omega)\in\cme)>0$.
 In particular, the results of Theorem \ref{thmintro2}, applied to
 the spin-fermion system, hold: the system approaches the repeated interaction asymptotic state $\varrho_+$,  defined in \fer{intro3'}.
\end{thm}

\noindent
{\bf Proof.\ } We expand the operator $M$ in a power (Dyson) series in $\lambda$:
\begin{eqnarray}
\lefteqn{M = \e^{\ri\tau L_\cS} P}\label{mgre38}\\
&&+ \sum_{n\geq 1} \lambda^{2n}\int_0^\tau \d t_1\cdots
\int_0^{t_{2n-1}} \d t_{2n} \  \e^{\ri \tau L_\cS} P \e^{\ri
t_{2n} K_0}W\e^{-\ri t_{2n}K_0} \cdots \e^{\ri t_1 K_0} W\e^{-\ri t_1
K_0} P, \nonumber
\end{eqnarray}
where only the even powers appear since the interaction is linear
in creation and annihilation operators, and $P$ projects onto the
vacuum. $W$ is the operator
\begin{equation}
\label{mmm1}
W =V - J\Delta^{1/2} V \Delta^{-1/2}J,
\end{equation}
where $V$ is given in \fer{interaction2} (see also
\fer{mgre35}), and $J=J_\cS\otimes
J_\cE,\Delta=\Delta_\cS\otimes\Delta_\cE$ are the modular
conjugation and the modular operator associated to
$(\fm_\cs\otimes\fm_\cE,\psi_\cS\otimes\psi_\cE)$, see also
\fer{mgre33}, \fer{chainmod1}. Using the Canonical Anticommutation
Relations, one easily sees that $\|a_\beta^{\#}(g)\|=\|g\|$
(independent of $\beta$; see \fer{mgre39} for the definition of
the thermal creation and annihilation operators). Using \fer{mmm1} and \fer{mgre35}, it is easy
to find the upper bound
\begin{equation}
\label{mgre40} \| W\| \leq 3 \|(1+\e^{\beta h/2})g\|.
\end{equation}
We apply standard analytic perturbation theory to the operator
\fer{mgre38}. For $\lambda=0$, the eigenvalues of $M$,
$\{1,\e^{\pm 2\ri\tau}\}$, lie apart by the distance
\begin{equation}\label{mgre41} 
r_0(\tau) := \min\left\{ 2|\sin(2\tau)|,
2|\sin(\tau)| \right\}.
\end{equation}
(Note the spectrum of $L_\cs$ is $\{-2,0,0,2\}$, c.f.
\fer{mgre32}.) We assume that the interaction time is such that
$r_0(\tau)$ is strictly positive. Below, this condition appears as
${\rm dist}(\tau,\frac{\pi}{2}{\mathbb N})>\delta>0$.

The following result gives an estimate of the eigenvalues of $M$,
which will be needed in verifying that $M$ is in the family
$\cme$.
%%%%%%%%%%%%%%%%%% PROP: SPIN FERMION EIGENVALUES %%%%%%%%%%%%%%%%%%%%%%%%%%%%
\begin{prop}\label{mgreprop1} 
Suppose that $|\lambda|<\frac 14 r_0(\tau)$.
Denote by $1,e_0,e_\pm$ the four eigenvalues of $M$. We have
\begin{eqnarray}
e_0 & = & 1-\lambda^2\tau^2 \alpha +\varepsilon_0,\label{mgre43}\\
e_\pm & = & \e^{\pm 2\ri\tau}\Big[ 1-\frac{\lambda^2\tau^2}{2}
\alpha \pm \ri\lambda^2\tau^2\int\d\mu(r) \|g(r)\|_{\frak g}^2
\times\nonumber\\
 & & \times \Big( \frac{1-\sinc(\tau(2-r))}{2-r} +
\frac{1-\sinc(\tau(2+r))}{2+r}\Big)\Big] +\varepsilon_\pm,
\label{mgre44}
\end{eqnarray}
where $\sinc(x)=\sin(x)/x$ and where
\begin{equation}
\label{mgre52}
 \alpha=  \int \d\mu(r) \|g(r)\|^2_{\frak g}
\Big(\sinc^2\big[\frac{\tau(r-2)}{2}\big]
+\sinc^2\big[\frac{\tau(r+2)}{2}\big] \Big).
\end{equation}
The error terms $\varepsilon_{\#}$, $\# = 0,\pm$, satisfy the bound
\begin{equation}\label{mgre37} 
|\varepsilon_{\#}|\leq 12 \lambda^4\tau^4 \|W\|^4
\cosh^2(|\lambda|\tau\|W\|)\Big[ 1+ \frac{1+\lambda^2
\tau^2\|W\|^2\cosh(|\lambda|\tau\|W\| )}{r_0(\tau)} \Big].
\end{equation}
\end{prop}
%%%%%%%%%%%%%%%%%%%%%%%%%%%%%%%%%%%%%%%%%%%%%%%%%%%%%%%%%%%%%%%%%%%%%%%%%%%%%%%%%

\proof Expansions
\fer{mgre43}, \fer{mgre44} of the eigenvalues have already been
calculated in \cite{bjm}, Section 4.8, but the error estimate
\fer{mgre37}, allowing the control of $\tau,\beta$, has not been
given there. This error estimate is obtained by performing
perturbation theory in a straightforward, but careful fashion. One
proceeds as in \cite{kato}, Chapter II.2. 
\qed

By knowing this expansion of the eigenvalues of $M$, we can impose
a smallness condition on $\lambda$ which guarantees that the
eigenvalues $e_{\#}$ have modulus strictly less than one, which is
equivalent to saying that $M\in\cme$.
%%%%%%%%%%%%%%% PROP: SPIN FERMION EIGENVALUE BOUNDS %%%%%%%%%%%%%%%%%%%%%%%%%%%%
\begin{prop}
\label{mgreprop3} Suppose that $\tau_{\rm min}<\tau <\tau_{\rm
max}$, $\beta<\beta_{\rm max}$, and that ${\rm
dist}(\tau,\frac{\pi}{2}{\mathbb N})>\delta$, for some constants
$0<\tau_{\rm min}<\tau_{\rm max}$ and $\beta_{\rm max}$, $\delta>0$.
Then there is a constant $\lambda_0>0$, depending on $\tau_{\rm
min}, \tau_{\rm max}, \beta_{\rm max}, \delta$, as well as on the
form factor $g$, s.t. if $0<|\lambda|<\lambda_0$, then
\begin{equation}
\label{mgre45} |e_{\#}| < 1-\frac{\lambda^2\tau^2}{8} \alpha <1,
\end{equation}
$\#=0,\pm$. In particular, $M\in\cme$.
\end{prop}
%%%%%%%%%%%%%%%%%%%%%%%%%%%%%%%%%%%%%%%%%%%%%%%%%%%%%%%%%%%%%%%%%%%%%%%%%%%%%%%%%

\noindent {\bf End of proof of Theorem \ref{mgrethm2}, given Proposition
\ref{mgreprop3}.\ } Fix $\tau_{\rm min}, \tau_{\rm max},
\beta_{\rm max}$ and $\delta$, and suppose that \fer{mgre36}
holds. Denote by $\Omega'$  the set of $\omega$ for which
 ${\rm dist}(\tau(\omega),\frac{\pi}{2}{\mathbb N})>\delta$. Then ${\rm
p}(\Omega')\neq 0$, and for each $\omega\in\Omega'$, we have
$M(\omega)\in\cme$, by Proposition \ref{mgreprop3}.  Consequently,
${\rm p}(M(\omega)\in\cme)\geq {\rm p}(\Omega')>0$. \qed

\noindent {\bf Proof of Proposition \ref{mgreprop3}.\ } We impose
conditions s.t. the three eigenvalues given in \fer{mgre43},
\fer{mgre44} have modulus strictly less than one. We have

\begin{equation}\label{1est} 
|e_0|< 1 - \frac{\lambda^2\tau^2}{2}\alpha,
\end{equation}
provided
\begin{equation}\label{1cond}
|\varepsilon_0|<\frac{\lambda^2\tau^2}{2}\alpha.
\end{equation}
Next, since $e_-$ is the complex conjugate of $e_+$, it suffices
to consider the latter. We write, with obvious identifications in
\fer{mgre44}, $e_+=\e^{2\ri\tau}[1-x+\ri y +\varepsilon_+]$. We
have 
\begin{equation}\label{mgre46} 
|e_+| \leq |1-x+\ri y| +|\varepsilon_+| = \sqrt{1-2x +x^2+y^2} +|\varepsilon_+|. 
\end{equation}
Since $x^2+y^2$ is just the square of the modulus of the second
order ($\lambda^2$) contribution to the eigenvalue $e_+$, it is
easy to see 
%(as in the proof of Proposition \ref{mgreprop1}) 
that $x^2+y^2\leq \lambda^4 \tau^4 \|W\|^4 (1+\frac{1}{r_0(\tau)})^2$.
We now impose the condition
\begin{equation}\label{mgre47} 
2\lambda^2\tau^2 \|W\|^4 \left(1+\frac{1}{r_0(\tau)}\right)^2 < \alpha,
\end{equation}
which implies that $x^2+y^2 < x$. Combining this latter inequality
with \fer{mgre46} gives
\begin{equation}\label{mgre49}
|e_+| < \sqrt{1-x} +|\varepsilon_+| \leq 1-x/2 + |\varepsilon_+|.
\end{equation}
Finally we impose the condition
\begin{equation}\label{mgre48} 
|\varepsilon_+| < x/4 = \frac{\lambda^2\tau^2}{8}\alpha,
\end{equation}
so that we get from \fer{mgre49}
\begin{equation}\label{mgre50}
|e_+|< 1-x/4 =  1-\frac{\lambda^2\tau^2}{8}\alpha. 
\end{equation}
This last bound, combined with \fer{1est}, proves that
\fer{mgre45} holds, provided the conditions \fer{mgre48},
\fer{mgre47} and \fer{1cond} are imposed. Taking into account the
bound \fer{mgre37}, we see that a sufficient condition for
\fer{mgre48}, \fer{mgre47} and \fer{1cond} to hold is that
\begin{equation}\label{mgre51} 
96\lambda^2\tau^2 \|W\|^4 \cosh^2(|\lambda|\tau\|W\|)\Big[ 1+ 
\frac{1+\lambda^2 \tau^2\|W\|^2\cosh(|\lambda|\tau\|W\| )}{r_0(\tau)} \Big] <
\alpha.
\end{equation}
One may now use \fer{mgre40}, \fer{mgre41}, to find a constant
$\lambda_0$, depending only on the parameters as stated in the
proposition, s.t. if  $|\lambda|<\lambda_0$, then \fer{mgre51}
holds. (Note that $\alpha$, \fer{mgre52}, does not depend on
$\beta$, and the minimum of $\alpha$, taken over $\tau>0$ varying in
any compact set, must be strictly positive.) This completes the
proof of Proposition \ref{mgreprop3}, and hence that of Theorem \ref{mgrethm2}. 
\qed

\subsection{Proof of Theorem \ref{thmintro3}}

Since $\E[M]\in\cme$ (by Theorems \ref{mgrethm2} and \ref{thmintro1}), $1$ is a simple eigenvalue of $\E[M]$. Let $\psi^*_{\cal S}$ denote the unique vector invariant under ${\mathbb E}[M^*]$, normalized as $\scalprod{\psi^*_{\cal S}}{\psi_{\cal S}}=1$, where $\psi_{\cal S}$ is given in \fer{mgre30}. We have $P_{1,{\mathbb E}[M]} = |\psi_{\cal S}\rangle\langle\psi^*_{\cal S}|$, and thus, by \fer{eq:thetalim}, $\theta=\psi^*_{\cal S}$. To calculate $\psi^*_{\cal S}$, we note first that for any $\omega$, $M(\omega)$ is block-diagonal:
\begin{lem}
\label{mlem1}
Let $P_0=|0\rangle\langle 0|+|1\rangle\langle 1|$ be the spectral projection of $L_{\cal S}$ associated to $\{1\}$. The operator $M(\omega)$ leaves the subspace ${\rm Ran}P_0$ invariant. In the ordered orthonormal basis $\{ |0\ket,|1\ket \}$ of ${\rm Ran}P_0$, we have the representation
\begin{equation}
\label{mmm2}
P_0 M(\omega) P_0 = \bbbone -\lambda^2\tau^2(\omega)
\left[
\begin{array}{cc}
\alpha_-(\omega) & -\alpha_-(\omega)\\
-\alpha_+(\omega) & \alpha_+(\omega)
\end{array}
\right] +O(\lambda^4),
\end{equation}
where the $\alpha_\pm(\omega)$ are given by \fer{mmm3} with $\tau(\omega)$ replaced by $\tau_0$. The remainder term is uniform in $\tau$ varying in compact sets.
\end{lem}

\noindent
{\bf Proof of Lemma \ref{mlem1}.\ } As explained at the beginning of the proof of Theorem \ref{mgrethm2}, only even powers of the interaction are present in the Dyson series expansion for $M$, \fer{mgre38}. It follows from \fer{interaction2} and \fer{mmm1} that each term in the Dyson series \fer{mgre38} leaves ${\rm Ran}P_0$ invariant; this is so because the operator $\sigma_x$ shows up an even number of times, and $\sigma_x |0\ket=|1\ket$ and $\sigma_x |1\ket=|0\ket$. The calculation of the explicit form \fer{mmm2} is not hard. This concludes the proof of Lemma \ref{mlem1}
\qed

\noindent
The expansion for $M^*$ and hence of ${\mathbb E}[M^*]$ in powers of $\lambda$ follow directly from \fer{mmm2}.  One then performs an expansion in powers of $\sigma$ and finds for the $O(\lambda^2)$-term: 
\begin{eqnarray*}
\lefteqn{ 
-\lambda^2 \tau^2_0 
\left[
\begin{array}{cc}
\alpha_- & -\alpha_+\\
-\alpha_- & \alpha_+
\end{array}
\right]
-
 2\lambda^2{\mathbb E}[\sigma]
\left[
\begin{array}{cc}
\xi_- & -\xi_+\\
-\xi_- & \xi_+
\end{array}
\right]
}\\
&& 
\qquad\qquad \qquad\qquad -\lambda^2 ({\mathbb E}[\sigma])^2
\left[
\begin{array}{cc}
\eta_- & -\eta_+\\ 
-\eta_- & \eta_+
\end{array}
\right]
+\lambda^2 O(\sigma^3).
\end{eqnarray*}
The following expansion of the invariant vector $\psi^*_{\cal S}$ follows:
\begin{eqnarray}
\frac{1}{\sqrt 2}\psi^*_{\cal S} &=& \frac{\alpha_+}{\alpha_++\alpha_-}|0\ket\otimes|0\ket +\frac{\alpha_-}{\alpha_++\alpha_-}|1\ket\otimes|1\ket\nonumber\\
&& +2{\mathbb E}[\sigma] \frac{\alpha_-\xi_+-\alpha_+\xi_-}{\tau_0^2(\alpha_++\alpha_-)^2}(|0\ket\otimes|0\ket - |1\ket\otimes |1\ket) \nonumber\\
&&+4({\mathbb E}[\sigma])^2(\xi_-+\xi_+)\frac{\alpha_-\xi_+-\alpha_+\xi_-}{\tau^4_0(\alpha_++\alpha_-)^3}(|0\ket\otimes |0\ket - |1\ket\otimes |1\ket)\nonumber\\
&&+{\mathbb E}[\sigma^2]\frac{\alpha_-\eta_+-\alpha_+\eta_-}{\tau_0^2(\alpha_++\alpha_-)^2} (|0\ket \otimes|0\ket - |1\ket\otimes |1\ket)\ + O(\sigma^3) +O(\lambda^2).\qquad
\label{mmm4}
\end{eqnarray}
Formula \fer{asstate} now follows directly from \fer{mmm4} and \fer{intro3'}. This concludes the proof of Theorem \ref{thmintro3}.\qed

%%%%%%%%%%%%%%%%%%%%%%%%%%%%%%%%%%%%%%%%%%%%%%%%%%%%%%%%%%%%%%%%%%%%%%%%%%%%%%%%%%%%%%%%%%%%%
%%%%%%%%%%%%%%%%%%%%%%%%%%%%%%%%%%%%%%%%%%%%%%%%%%%%%%%%%%%%%%%%%%%%%%%%%%%%%%%%%%%%%%%%%%%%%
%%%%%%%%%%%%%%%%%%%%%%%%%%%%%%%%%%%%%%%%%%%%%%%%%%%%%%%%%%%%%%%%%%%%%%%%%%%%%%%%%%%%%%%%%%%%%
%%%%%%%%%%%%%%%%%%%%%%%%%%%%%%%%%%%%%%%%%%%%%%%%%%%%%%%%%%%%%%%%%%%%%%%%%%%%%%%%%%%%%%%%%%%%%


\begin{thebibliography}{}
\bibitem{AW} Araki, H.,  Wyss, W., ``Representations of canonical anticommutation relations'', {\em Helv. Phys. Acta} {\bf 37}, 136-159 (1964). 

%\bibitem{la} Arnold, L., Random Dynamical Systems, Springer, 2003.

\bibitem{AJ} Attal, S., Joye, A., ``The Langevin Equation for a Quantum Heat Bath'' ,
{\em J. Func. Anal.} {\bf 247}, 253-288 (2007).

\bibitem{AJP}
Attal, S., Joye, A., Pillet, C.-A. (Eds.), Open Quantum Systems
I-III, Lecture Notes in Mathematics, volumes 1880-1882, Springer
Verlag, 2006.

\bibitem{az} Azoff, E.A., ``Borel Measurability in Linear Algebra'',
{\em Proc. Am. Math. Soc.} {\bf 42} (2), 346-350 (1974).

%\bibitem{beckschwarz} Beck, A., Schwartz, J.T., ``A Vector-Valued Random Ergodic
%Theorem'', {\em Proc. Am. Math. Soc.}, {\bf 8}, no.6, 1049-1059 (1957).

\bibitem{BR}
Bratteli, O., Robinson, D.W., Operator Algebras and Quantum
Statistical Mechanics, volumes 1 and 2, Texts and Monographs in
Physics, Springer Verlag, 1996.

%\bibitem{ben} Bru, R., Elsner, L., Neumann, M., ``Convergence of Infinite Products
%of Matrices and Inner-Outer Iteration Schemes'', {\em
%Elect. Trans. Num. Anal.}, {\bf 2}, 183-193 (1994).

\bibitem{B} Bruneau, L., ``Repeated interaction quantum systems'', Proceedings of the IRS conference 2007, to appear in Markov Process. Related Fields.

\bibitem{bjm} Bruneau, L., Joye, A., Merkli, M., ``Asymptotics of repeated interaction quantum systems'', {\em J. Func. Anal.} {\bf 239}, 310-344 (2006).

\bibitem{bjm2} Bruneau, L., Joye, A., Merkli, M., ``Infinite products of random matrices and repeated interactions dynmaics'', preprint arxive:math.PR/0703625.

%\bibitem{g} Guivarc'h, Y., ``Limit Theorem for Random Walks and Products of RandOm Matrices'', in the {\em Proceedings of the CIMPA-TIFR School on
%Probability measures on Groups, Recent Directions and Trends, sept. 2002}, TIFR Mumbai.

%\bibitem{hoef} Hoeffding, W., ``Probability inequalities for sums of
%bounded random variables'', {\em Journal of the American Statistical
%Association} {\bf 58}, 13-30 (1963).

\bibitem{jp2002} Jak\u{s}ic, V., Pillet, C.-A., ``Non-equilibrium steady states of
finite quantum systems coupled to thermal reservoirs'', {\em
Commun. Math. Phys.} {\bf 226}, 131-162 (2002).

\bibitem{JP1}
Jak\u{s}i\'c, V., Pillet, C.-A.: {\it A note on the entropy
production formula.}  Advances in differential equations and
mathematical physics (Birmingham, AL, 2002),  175--180, Contemp.
Math., 327, Amer. Math. Soc., Providence, RI, 2003.


\bibitem{kato} Kato, K., Perturbation Theory for Linear Operators. $2nd$
      edition. Springer, Berlin, 1976.

%\bibitem{ks} Kesten, H., Spitzer, F., ``Convergence in Distribution of
%Products of Random Matrices'', {\em Z. Wahrscheinlichkeitstheorie verw.
%Gebiete}, {\bf 67}, 363-386 (1984).

%\bibitem{k} Kifer, Y., Liu, P.-D., Random Dynamics, in {\em Handbook of
%Dynamical Systems}, vol1B, Hasselblatt and Katok Edts, 379-499,
%North-Holland, 2006.

%\bibitem{M} Merkli, M., Proceedings IAMP (2007).

\bibitem{MMS}
Merkli, M., M\"uck, M., Sigal, I.M., ``Instability of Equilibrium
States for Coupled Heat Reservoirs at Different Temperatures'',
{\em J. Funct. Anal.} {\bf  243}, 87-120 (2007).

\bibitem{MMS2}
Merkli, M., M\"uck, M., Sigal, I.M., ``Theory of Non-Equilibrium
Stationary Sates as a Theory of Resonances", to appear in {\em
Ann. H. Poincar\'e}, 2007.

\bibitem{MSB}
Merkli, M., Sigal, I.M., Berman, G.P., "Resonance Theory of
Decoherence and Thermalization", to appear in {\em Ann. Phys.},
2007; and ``Decoherence and thermalization'', {\em Phys. Rev. Lett.} {\bf  98}, no.13, 130401 (2007).

\bibitem{MWM}
Meschede, D., Walther, H., M\"uller, G., ``One-atom maser'', {\em
Phys. Rev. Lett.} {\bf 54}, 551-554 (1993).

\bibitem{WVHW}
Weidinger, M., Varcoe, B.T.H., Heerlein, R., Walther, H., ``Trapping
states in micromaser'', {\em Phys. Rev. Lett.} {\bf 82}, 3795-3798
(1999).

\bibitem{WBKM}
Wellens, T., Buchleitner, A., K\"ummerer, B., Maassen, H., ``Quantum
state preparation via asymptotic completeness'', {\em Phys. Rev.
Lett.} {\bf 85}, 3391-3364 (2000).


\end{thebibliography}
\end{document}